\documentclass[aps,pre,preprint,showpacs,eqsecnum,amssymb]{revtex4}
\usepackage{graphicx,color,psfrag}
\begin{document}
%\date{\today} t
\title{Aspects of the disordered harmonic chain}
\author{Hans C. Fogedby}
\email{fogedby@phys.au.dk}
\affiliation{Department of Physics and
Astronomy, University of
Aarhus, Ny Munkegade\\
8000 Aarhus C, Denmark\\}
\begin{abstract}
We discuss the driven harmonic chain with fixed boundary conditions
subject to weak coupling strength disorder. We discuss the evaluation of the 
Liapunov exponent in some detail expanding on the dynamical system theory
approach by Levi et al. We show that including mass disorder the mass and
coupling strength disorder can be combined in a renormalised mass disorder. 
We review the method of Dhar regarding the disorder-averaged heat current,
apply the approach to the disorder-averaged large deviation function and finally 
comment on the validity of the Gallavotti-Cohen fluctuation theorem.
The paper is also intended as an introduction to the field and includes
detailed calculations.
\end{abstract}
.

\maketitle
%%%%%%%%%%%%%%%%%%%%%%%%%%%%%%%%%%%%%%%%%%%%%%%%%%%%%%%%%
%%%%%%%%%%%%%%%%%%%%%%%%%%%%%%%%%%%%%%%%%%%%%%%%%%%%%%%%%
%%%%%%%%%%%%%%%%%%%%%%%%%%%%%%%%%%%%%%%%%%%%%%%%%%%%%%%%%
\section{\label{intro} Introduction}
%%%%%%%%%%%%%%%%%%%%%%%%%%%%%%%%%%%%%%%%%%%%%%%%%%%%%%%%%
%%%%%%%%%%%%%%%%%%%%%%%%%%%%%%%%%%%%%%%%%%%%%%%%%%%%%%%%%
%%%%%%%%%%%%%%%%%%%%%%%%%%%%%%%%%%%%%%%%%%%%%%%%%%%%%%%%%
There is a current interest in small fluctuating systems in contact
with heat reservoirs driven by external forces. This focus is driven
by the recent possibilities of direct manipulation of nano systems 
and bio molecules. These techniques also permit direct experimental
access to the probability distributions for the work or heat exchanged 
with the environment  \cite{Trepagnier04,Collin05,Seifert06a,Seifert06b,
Imparato07,Ciliberto06,Ciliberto07, Ciliberto08}. These single molecule techniques
have, moreover, also yielded access to the so-called fluctuation theorems,
which relate the probability of observing entropy-generated trajectories, 
with that of observing entropy-consuming trajectories
\cite{Jarzynski97,Kurchan98,Gallavotti96,Crooks99,Crooks00,
Seifert05a,Seifert05b,Evans93,Evans94,Gallavotti95,
Lebowitz99,Gaspard04,Imparato06,
vanZon03,vanZon04,vanZon03a,vanZon04a,Seifert05c}.

A fundamental issue is the validity and microscopic underpinning of Fourier's law  \cite{Bonetto00,Jackson78}.
Here an important problem is the dependence of the heat current $J$ on the system size $N$
and dimensionality.  Fourier's law based on energy conservation and a phenomenological transport equation
assumes local equilibrium and therefore a current $J\sim 1/N$, yielding a constant heat conductivity 
$\kappa\propto JN$. However, many studies of one dimensional systems  indicate that  $J\sim 1/N^\alpha$, 
where $\alpha$ in general is different from one, signalling the breakdown of Fourier's law, 
see \cite{Dhar08a,Lepri03,Dhar16}. Regarding ongoing studies of the dependence of $J(N)$ as function of boundary conditions and the spectral properties of the heat baths, see e.g. \cite{Ajanki11,Ash20,Ong14,Yamada18,Amir18,Herrera10,Herrera15,Herrera19,Zhou16,Kundu10a}. 

A one dimensional system which has been studied extensively is the linear harmonic chain subject to disorder or 
nonlinearity. In the case of a linear harmonic chain the heat is transmitted ballistically by phonons and the heat 
current is independent of the system size, corresponding to $\alpha=0$ \cite{Saito11,Kundu11,Fogedby12}. 
In the particular case where the effective interaction is provided by mass disorder, this issue has been studied
in several papers 
\cite{Casher71,Dhar08a,Dhar08b,Dhar11,Kundu10,Chauduri10,Lee05,Lepri03,Oconnor74,Roy08a,Roy08b}. 
For more recent papers on the disordered chain, see also 
\cite{Herrera10,Herrera19,Ash20,Amir18,Ong14,Ajanki11,Kundu10a,Yamada18}.

The status regarding the mass-disordered chain has been summarised by Dhar \cite{Dhar08a}, 
see also Lepri \cite{Lepri03}. Unlike the electronic case, where disorder gives rise to Anderson localisation 
\cite{Anderson58} of the carriers and thus a vanishing contribution to the current, the case of phonons subject 
to disorder is different. Translational invariance implies that the low frequency phonon modes are extended and 
thus contribute to the current \cite{Matsuda62,Matsuda70,Ishii73}. Moreover, unlike the electronic 
case where only electrons at the Fermi surface contribute, the phonon contributions originate from the full phonon 
band. At larger frequencies corresponding to larger wave numbers, i.e., smaller wavelengths, the disorder becomes 
effective and traps the phonons in localised states. As a result the high frequency localised phonons 
mode do not contribute to the heat current. 

For an ordered linear chain, where the heat is carried ballistically across the system by extended phonon
modes, the heat current and more generally the large deviation function monitoring the heat
fluctuations are easily evaluated explicitly by means of standard techniques \cite{Saito07,Kundu11,Fogedby12}. 
On the other hand, for a disordered system standard techniques using plane wave representations fail and one
must resort to transfer matrix methods in order to monitor and analyse the propagation of lattice site vibrations
 \cite{Matsuda70,Ishii73,Dhar01}. 

In the mass-disordered case detailed analysis by Matsuda et al.  \cite{Matsuda70,Ishii73} based on the 
Furstenberg's theorem for the product of random matrices \cite{Furstenberg63,O'Connor75} yield 
a Liapunov exponent $\gamma(\omega)$ depending on the phonon frequency $\omega$. Regarding the
definition of the Liapunov exponent, consider the two-by-two random matrix $T_n(\omega)$ relating the pair of
displacements $u_n,u_{n-1}$ to the displacements $u_{n+1}, u_n$. Here the Liapunov exponent
basically characterises the growth of an ordered product of statistically independent random matrices 
according to $\prod_{n=1}^N T_n\sim\exp(\gamma(\omega)N)$, or more precisely
$(|u_N|^2+|u_{N-1}|^2)^{1/2}\sim\exp(\gamma(\omega) N)$. From the analysis of Matsuda et al. it follows
that $\gamma(\omega)\propto\omega^2$ for small $\omega$ and we infer a localisation length $l_c(\omega)=1/\gamma(\omega)$ and in particular a  cross-over 
frequency  $\omega_c=\text{A}N^{-1/2}$, where $A$ depends on the disorder; for an ordered chain $A=0$. 
Consequently, the high frequency phonons for $\omega>\omega_c$, corresponding to small wavelengths, are 
trapped and do not contribute to the transport, whereas low frequency phonons with $\omega<\omega_c$ carry 
the heat across the chain. The dependence of $\omega_c$ on the system size $N$ implies a dependence of the
scaling exponent $\alpha$ in $J\sim 1/N^\alpha$. Moreover, boundary conditions and spectral properties of
the heat reservoirs also influence $\alpha$ \cite{Dhar01}.  Assuming that the real part of the frequency dependent
 damping, $\Gamma'(\omega)\sim |\omega|^s$, one finds $\alpha=3/2+s/2$, the case of an unstructured reservoir,
$\Gamma=\text{const.}$, $s=0$ , yields $\alpha=3/2$; note that $s=-1$ results in $\alpha=1$, i.e., Fourier's law.

In the present paper we consider a harmonic chain with fixed boundary conditions subject
to weak coupling strength disorder. To our knowledge this case has not been discussed previously.
Although the techniques used by Matsuda et al. in their pioneering work in the case of mass disorder
\cite{Matsuda70,Ishii73} presumably can be applied to the case of coupling strength disorder, we have 
found the approach in the mass-disordered case by Lepri et al. \cite{Lepri03} using dynamical system theory 
in conjunction with statistical physics more accessible and transparent. The main part of the paper 
is thus devoted to the evaluation and discussion of the Liapunov exponent, a central quantity, in the case of 
coupling strength disorder. We have, moreover, briefly discussed the heat current and the heat fluctuations
characterised by the large deviation function. In order to render the paper self-contained for the uninitiated reader, 
we have presented a review of the methods employed including a series of explicit calculations. These calculations 
are deferred to a series of appendices.

The paper is organised as follows. In Sec.~\ref{model} we introduce the disordered harmonic chain 
in the presence of both mass disorder and coupling strength disorder, driven at the end points by heat 
reservoirs, the equations of motion describing the dynamics together with expressions for the heat rates. 
In Sec.~\ref{green} we present the Green's functions describing the propagation of phonons for the ordered 
and disordered chain with derivations deferred to Appendix \ref{greensfunction}. In Sec.~\ref{liapunov} we 
introduce the Liapunov exponent characterising the asymptotics of the transfer matrices. 
Sec.~\ref{disorder}, which is the central part of the paper, is devoted to a derivation of the Liapunov exponent 
in the case of weak coupling strength disorder with a technical issue deferred to Appendix \ref{liapunovapp}.
In Sec.~\ref{heatldf} we briefly discuss the heat current and the large deviation function with derivations in
Appendices \ref{current} and \ref{cgf}. In Sec.~\ref{conclusion} 
we present a conclusion. 
%%%%%%%%%%%%%%%%%%%%%%%%%%%%%%%%%%%%%%%%%%%%%%%%%%%%%%%%%
%%%%%%%%%%%%%%%%%%%%%%%%%%%%%%%%%%%%%%%%%%%%%%%%%%%%%%%%%
%%%%%%%%%%%%%%%%%%%%%%%%%%%%%%%%%%%%%%%%%%%%%%%%%%%%%%%%%
\section{\label{model} Model}
\begin{figure}
\centering
\includegraphics[width=1.0\hsize]{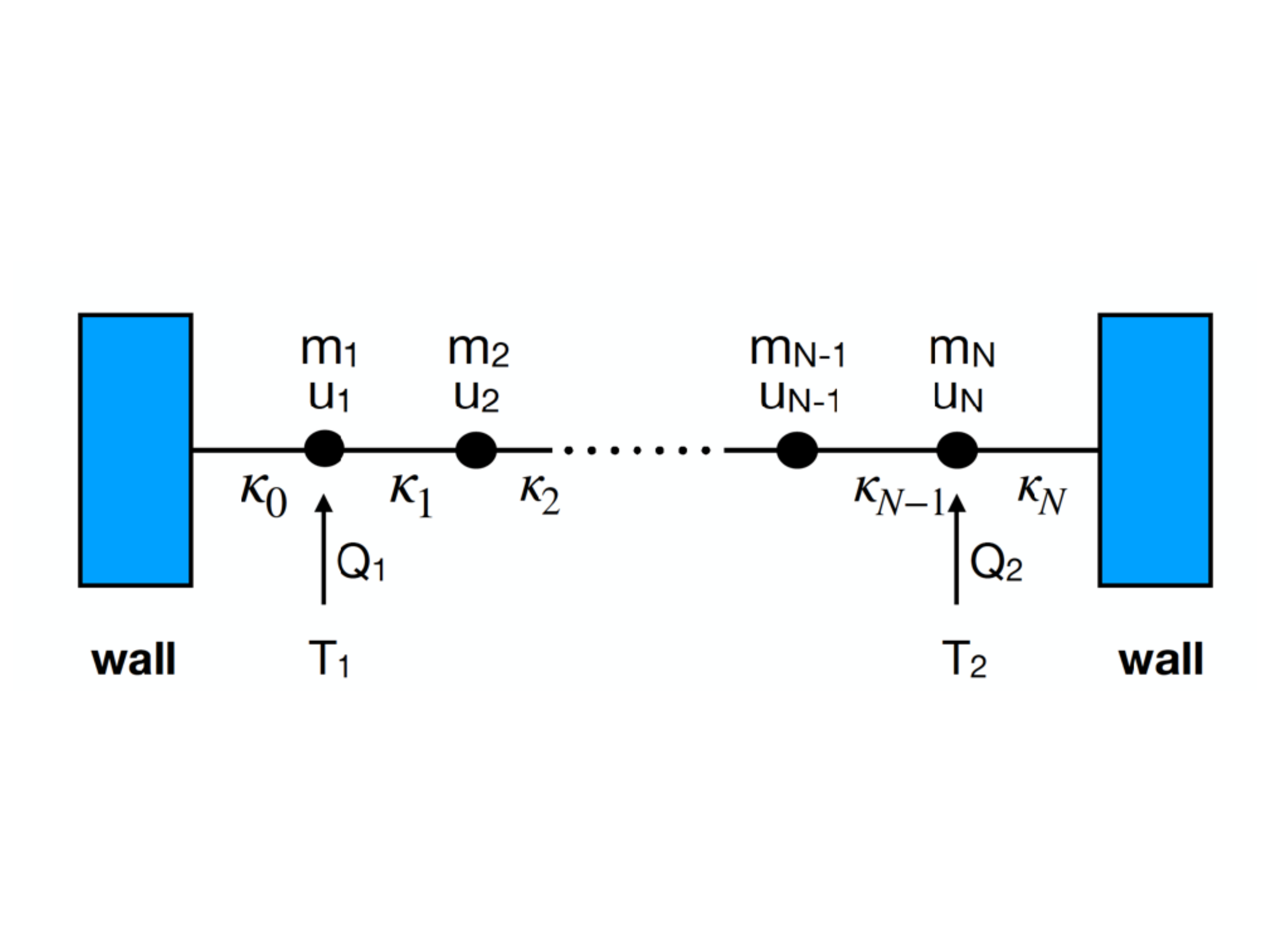}
\caption{Depiction of the random mass and  coupling strength harmonic chain. The particle positions are
denoted by $u_n$, the masses by $m_n$, and the coupling strengths by $\kappa_n$. The particles at 
$n=1$ and $n=N$ are attached to the walls. The chain is driven by heat reservoirs at $n=1$ and $n=N$ 
transmitting the heat $Q_1$ and $Q_2$, respectively. The heat reservoirs are maintained at temperatures 
$T_1$ and $T_2$, respectively.}
\label{fig1}
\end{figure}

%%%%%%%%%%%%%%%%%%%%%%%%%%%%%%%%%%%%%%%%%%%%%%%%%%%%%%%%%
%%%%%%%%%%%%%%%%%%%%%%%%%%%%%%%%%%%%%%%%%%%%%%%%%%%%%%%%%
%%%%%%%%%%%%%%%%%%%%%%%%%%%%%%%%%%%%%%%%%%%%%%%%%%%%%%%%%
Here we introduce the model which is the subject of the present study.
The disordered harmonic chain of length $N$ is characterised by the Hamiltonian
\begin{eqnarray}
H=\sum_{n=1}^N\frac{\dot u_n^2}{2m_n}+\frac{1}{2}\sum_{n=1}^{N-1}\kappa_n(u_n-u_{n+1})^2
+\frac{1}{2}\kappa_0 u_1^2+\frac{1}{2}\kappa_Nu_N^2,
\label{ham}
\end{eqnarray}
where $u_n$ is the position of the n-th particle and $\dot u_n$ its velocity, $\dot u_n=du_n/dt$. 
We consider fixed boundary conditions, i.e., the chain is attached to walls at the endpoints, yielding the 
on-site potentials $(1/2)\kappa_0u_1^2$ and $(1/2)\kappa_Nu_N^2$. For later purposes we impose both mass 
disorder and coupling strength disorder, i.e., $m_n$ and $\kappa_n$ are determined by the independent distributions
$\pi_1(m_n)$ and $\pi_2(\kappa_n)$. 
The chain is driven by two reservoirs at temperatures $T_1$ and $T_2$, respectively, acting on the first and 
last particle in the chain. This configuration is depicted in Fig.~\ref{fig1}.

The equations of motion in bulk for $1<n<N$ and the Langevin equations for the endpoints 
for $n=1$ and $n=N$ are given by 
\begin{eqnarray}
&&m_n\ddot u_n(t)=\kappa_nu_{n+1}(t)+\kappa_{n-1}u_{n-1}(t)-(\kappa_n+\kappa_{n-1})u_n(t),
~~1<n<N,
\label{eqmo1}
\\
&&m_1\ddot u_1(t)=\kappa_1u_2(t)-(\kappa_1+\kappa_0)u_1(t)-\Gamma\dot u_1(t)+\xi_1(t),
\label{lan1}
\\
&&m_N\ddot u_N(t)=\kappa_{N-1}u_{N-1}(t)-(\kappa_N+\kappa_{N-1})u_N(t)-\Gamma\dot u_N(t)+\xi_2(t).
\label{lan2}
\end{eqnarray}
The heat reservoirs at the endpoints are characterised by 
the white noise correlations
\begin{eqnarray}
&&\langle\xi_1(t)\xi_1(t')\rangle=2\Gamma T_1\delta(t-t'),
\label{noi1}
\\
&&\langle\xi_2(t)\xi_2(t')\rangle= 2\Gamma T_2\delta(t-t');
\label{noi2}
\end{eqnarray}
note that the fluctuation-dissipation theorem  \cite{Reichl98} implies that the damping $\Gamma$
in the Langevin equations is balanced by the damping $\Gamma$ also appearing in the white noise correlations. 
We, moreover, consider  structureless reservoirs characterised by a single damping constant $\Gamma$. 
The case of memory effects characterised by a frequency dependent damping $\Gamma(\omega)$ has
also been discussed, see e.g. \cite{Dhar01,Casher71}. For later purposes we also note that according 
to (\ref{lan1}) and (\ref{lan2}) the thermal forces arising from the reservoirs are given by 
$F_1(t)=-\Gamma\dot u_1(t)+\xi_1(t)$ and $F_2(t)=-\Gamma\dot u_N(t)+\xi_2(t)$, yielding the heat rates
\begin{eqnarray}
&&\dot Q_1(t)=F_1(t)\dot u_1(t),
\label{qr1}
\\
&&\dot Q_2(t)=F_2(t)\dot u_N(t).
\label{qr2}
\end{eqnarray}
%
%%%%%%%%%%%%%%%%%%%%%%%%%%%%%%%%%%%%%%%%%%%%%%%%%%%%%%%%%
%%%%%%%%%%%%%%%%%%%%%%%%%%%%%%%%%%%%%%%%%%%%%%%%%%%%%%%%%
%%%%%%%%%%%%%%%%%%%%%%%%%%%%%%%%%%%%%%%%%%%%%%%%%%%%%%%%%
\section{\label{green} Green's functions}
%%%%%%%%%%%%%%%%%%%%%%%%%%%%%%%%%%%%%%%%%%%%%%%%%%%%%%%%%
%%%%%%%%%%%%%%%%%%%%%%%%%%%%%%%%%%%%%%%%%%%%%%%%%%%%%%%%%
%%%%%%%%%%%%%%%%%%%%%%%%%%%%%%%%%%%%%%%%%%%%%%%%%%%%%%%%%
The Green's function plays an important role in the discussion of heat transport and heat fluctuations. Introducing 
the Fourier transforms  
\begin{eqnarray}
&&u_n(t)=\int\frac{d\omega}{2\pi}\exp(-i\omega t)\tilde u_n(\omega),
\label{fou1}
\\
&&\xi_{1,2}(t)=\int\frac{d\omega}{2\pi}\exp(-i\omega t)\tilde\xi_{1,2}(\omega),
\label{fou2}
\end{eqnarray}
we can express the equations of motion (\ref{eqmo1}) to (\ref{lan2}) in the form
\begin{eqnarray}
\sum_{m=1}^NG_{nm}^{-1}(\omega)\tilde u_m(\omega)=\delta_{n1}{\tilde\xi}_1(\omega)+\delta_{nN}{\tilde\xi}_2(\omega),
\label{eqmo2}
\end{eqnarray}
with solutions
\begin{eqnarray}
\tilde u_n(\omega)=G_{n1}(\omega)\tilde\xi_1(\omega)+G_{nN}(\omega)\tilde\xi_2(\omega),
\label{sol}
\end{eqnarray}
where the Green's function $G_{n1}(\omega)$ and $G_{nN}(\omega)$ describe the influence of the coupling to the 
reservoirs at the endpoints on the particle at site $n$. Here the end-to-end-point Green's function $G_{1N}(\omega)$ 
is relevant in the context of heat transfer.
%%%%%%%%%%%%%%%%%%%%%%%%%%%%%%%%%%%%%%%%%%%%%%%%%%%%%%%%%
%%%%%%%%%%%%%%%%%%%%%%%%%%%%%%%%%%%%%%%%%%%%%%%%%%%%%%%%%
\subsubsection{The ordered chain}
%%%%%%%%%%%%%%%%%%%%%%%%%%%%%%%%%%%%%%%%%%%%%%%%%%%%%%%%%
%%%%%%%%%%%%%%%%%%%%%%%%%%%%%%%%%%%%%%%%%%%%%%%%%%%%%%%%%
For the ordered chain with masses $m_n=m$ and coupling strengths $\kappa_n=\kappa$ the derivation
of $G_{1N}(\omega)$ is straightforward in a plane wave basis using an equation of motion approach
\cite{Fogedby12} or a determinantal approach \cite{Saito07,Kundu11}. For a chain composed of $N$ particles 
one finds the expression
\begin{eqnarray}
&&G_{1N}(\omega)=
\frac{\kappa\sin p}
{\kappa^2\sin p(N+1)-2i\kappa\Gamma\omega\sin pN-(\Gamma\omega)^2\sin p(N-1)},
\label{green1}
\\
&&\omega^2=\frac{4\kappa}{m}\sin^2(p/2),~~-\pi<p<\pi. 
\label{disp}
\end{eqnarray}
The denominator in (\ref{green1}) shows the resonance structure in the chain. We note that $G_{1N}(\omega)$ is
bounded and describes the propagation of ballistic phonons across the chain. The frequency $\omega$ is related 
to the wavenumber $p$ by the phonon dispersion law (\ref{disp}). The derivation of (\ref{green1}) is presented in 
Appendix  \ref{greensfunction}. 
%%%%%%%%%%%%%%%%%%%%%%%%%%%%%%%%%%%%%%%%%%%%%%%%%%%%%%%%%
%%%%%%%%%%%%%%%%%%%%%%%%%%%%%%%%%%%%%%%%%%%%%%%%%%%%%%%%%
\subsubsection{The disordered chain}
%%%%%%%%%%%%%%%%%%%%%%%%%%%%%%%%%%%%%%%%%%%%%%%%%%%%%%%%%
%%%%%%%%%%%%%%%%%%%%%%%%%%%%%%%%%%%%%%%%%%%%%%%%%%%%%%%%%
The  mass-disordered chain has been discussed by Dhar \cite{Dhar01,Dhar08a}, see also
\cite{Matsuda68,Matsuda70,Ishii73}. In this context the corresponding end-to-end Greens function $G_{1N}(\omega)$ 
has been derived. Here we extend this analysis to also include coupling strength disorder. For the disordered chain 
the plane wave assumption used in obtaining (\ref{green1}) is not applicable owing to the random masses and coupling 
strengths  and one must resort to a transfer matrix method \cite{Matsuda70,Ishii73,Dhar01}.

The transfer matrix $T_n(\omega)$ connects the pair of sites $(\tilde u_n(\omega),\tilde u_{n-1}(\omega))$ to the pair 
of sites $(\tilde u_{n+1}(\omega),\tilde u_n(\omega))$ and thus depends on the local disorder. From the bulk equations 
of motion (\ref{eqmo1}) in Fourier space we obtain
\begin{eqnarray}
\left(\begin{array}{c}\tilde u_{n+1}(\omega)\\\tilde u_n(\omega)\end{array}\right)=
T_n(\omega)\left(\begin{array}{c}\tilde u_n(\omega)\\\tilde u_{n-1}(\omega)\end{array}\right),
\label{eqmo3}
\end{eqnarray}
where the transfer matrix is given by
\begin{eqnarray}
T_n(\omega)=\left(\begin{array}{cc}\Omega_n(\omega)/\kappa_n &-\kappa_{n-1}/\kappa_n\\1 & 0\end{array}\right);
\label{trans1}
\end{eqnarray}
we have introduced
\begin{eqnarray}
\Omega_n(\omega)=\kappa_n+\kappa_{n-1}-m_n\omega^2.
\label{om}
\end{eqnarray}
The pair of sites $(\tilde u_N(\omega),\tilde u_{N-1}(\omega))$ are thus related to the pair of sites  
$(\tilde u_2(\omega),\tilde u_1(\omega))$ by a product of random transfer matrices according to
\begin{eqnarray}
\left(\begin{array}{c}\tilde u_N(\omega)\\\tilde u_{N-1}(\omega)\end{array}\right)=
T_{N-1}(\omega)T_{N-2}(\omega)\cdots T_2(\omega)\left(\begin{array}{c}\tilde u_2(\omega)\\\tilde u_1(\omega)\end{array}\right).
\label{eqmo4}
\end{eqnarray}
Incorporating the coupling to the heat baths at sites $n=1$ and  $n=N$  the Green's function 
$G_{1N}(\omega)$ takes the form 
\begin{eqnarray}
G_{1N}(\omega)=\frac{\kappa_0}
{\kappa_0\kappa_NB_{11}(\omega)+i\Gamma\omega(\kappa_NB_{12}
(\omega)-\kappa_0B_{21}(\omega))+(\Gamma\omega)^2B_{22}(\omega)}.
\label{green2}
\end{eqnarray}
Here $B(\omega)$ is given by the matrix product
\begin{eqnarray}
B(\omega)=T_N(\omega)T_{N-1}(\omega)\cdots T_1(\omega);
\label{bom}
\end{eqnarray}
note that in the ordered chain, $m_n=m$ and $\kappa_n=\kappa$, and we have $T_n(\omega)=T(\omega)$,
i.e., $B(\omega)=T^N(\omega)$. By insertion of $T^N(\omega)$ given by (\ref{a35}) we readily obtain (\ref{green1}).
The derivation of (\ref{green2}) is given in Appendix  \ref{greensfunction}. 
%%%%%%%%%%%%%%%%%%%%%%%%%%%%%%%%%%%%%%%%%%%%%%%%%%%%%%%%%
%%%%%%%%%%%%%%%%%%%%%%%%%%%%%%%%%%%%%%%%%%%%%%%%%%%%%%%%%
%%%%%%%%%%%%%%%%%%%%%%%%%%%%%%%%%%%%%%%%%%%%%%%%%%%%%%%%%
\section{\label{liapunov}The Liapunov exponent}
%%%%%%%%%%%%%%%%%%%%%%%%%%%%%%%%%%%%%%%%%%%%%%%%%%%%%%%%%
%%%%%%%%%%%%%%%%%%%%%%%%%%%%%%%%%%%%%%%%%%%%%%%%%%%%%%%%%
%%%%%%%%%%%%%%%%%%%%%%%%%%%%%%%%%%%%%%%%%%%%%%%%%%%%%%%%
The Liapunov exponent is of  importance in determining the properties of the disordered chain.
For large $N$ the behaviour of $G_{1N}(\omega)$ in (\ref{green2}) is determined by the asymptotic properties of the 
matrix product $B(\omega)$ in (\ref{bom}) . This issue has been discussed extensively in seminal papers by 
Matsuda and Ishii \cite{Matsuda68,Matsuda70,Ishii73} on the basis of the Furstenberg theorem 
\cite{Furstenberg63,O'Connor75}.  A central result is that
\begin{eqnarray}
\lim_{N\rightarrow\infty}\frac{1}{N}\log(|\tilde u_N(\omega)|^2+|\tilde u_{N-1}(\omega)|^2)=2\gamma(\omega)
\label{liap1}
\end{eqnarray}
with probability one;  this is basically an expression of the law of large numbers applied to non commuting
independent random matrices. 
Here $\gamma(\omega)$ is a positive Liapunov exponent depending on the phonon frequency 
$\omega$ and the disorder. For the norm we infer the scaling behaviour 
\begin{eqnarray}
|\tilde u_N(\omega)|\propto\exp(\gamma(\omega)N),
\label{un}
\end{eqnarray}
and since the matrix $T_N^{-1}(\omega)B(\omega)T_1^{-1}(\omega)$ according to (\ref{eqmo4}) 
connects the pair of site $(\tilde u_{N}(\omega),\tilde u_{N-1}(\omega))$ to the pair of sites 
$(\tilde u_{2}(\omega),\tilde u_{1}(\omega))$, the matrix elements $B_{nm}(\omega)$, likewise, scale like
\begin{eqnarray}
|B_{nm}(\omega)|\propto\exp(\gamma(\omega)N)
\label{bom2}
\end{eqnarray}
for large $N$. For vanishing disorder $\gamma(\omega)=0$ and $B(\omega)$ is
bounded, i.e., not growing with $N$. It then follows from (\ref{green2}) that  $G_{1N}(\omega)$, likewise,
is bounded.

Introducing the ratio $z_n(\omega)=\tilde u_n(\omega)/\tilde u_{n-1}(\omega)$ and inserting (\ref{om}), it follows 
from the bulk equations of motion (\ref{eqmo1}) that $z_n(\omega)$ obeys the non linear stochastic discrete map
\begin{eqnarray}
z_{n+1}(\omega)=
\frac{\kappa_n+\kappa_{n-1}-m_n\omega^2}{\kappa_n}-\frac{\kappa_{n-1}}{\kappa_n}\frac{1}{z_n(\omega)}.
\label{map1}
\end{eqnarray}
We also note from (\ref{un}) that for large $N$ we have 
$|z_N(\omega)|=|\tilde u_N(\omega)|/|\tilde u_{N-1}(\omega)|\propto\exp(\gamma(\omega))$ or
\begin{eqnarray}
\gamma(\omega)=\lim_{N\rightarrow\infty}\log| z_N(\omega)|.
\label{liap2}
\end{eqnarray}
Consequently, the Liapunov exponent $\gamma(\omega)$ is determined by the asymptotic properties of the map 
(\ref{map1}) for large $N$.

More precisely, in general the map (\ref{map1}) is stochastic due to the randomness of $m_n$ and $\kappa_n$.
However, just as a white Gaussian noise $\xi(t)$ with correlations $\langle\xi(t)\xi(t')\rangle=2\Delta\delta(t-t')$ in a 
Langevin equation of the form $dx(t)/dt=-dV(x)/dx+\xi(t)$ for a stochastic variable $x(t)$ can drive $x$ into a stationary 
distribution $P_0(x)\propto\exp(-V/\Delta)$ \cite{Reichl98}, we anticipate that the 'noise' due to the randomness of $m$ 
and $\kappa$ will drive $z_n$ into a stationary distribution $P_0(z)$. Consequently, according to (\ref{liap2}) we infer
\begin{eqnarray}
\gamma(\omega)=\int dzP_0(z)\log| z(\omega)|.
\label{liap3}
\end{eqnarray}
The task is thus to determine $P_0(z)$ on the basis of the map (\ref{map1}) and evaluate $\gamma(\omega)$.
%%%%%%%%%%%%%%%%%%%%%%%%%%%%%%%%%%%%%%%%%%%%%%%%%%%%%%%%%
%%%%%%%%%%%%%%%%%%%%%%%%%%%%%%%%%%%%%%%%%%%%%%%%%%%%%%%%%
%%%%%%%%%%%%%%%%%%%%%%%%%%%%%%%%%%%%%%%%%%%%%%%%%%%%%%%%%
\section{\label{disorder}Coupling strength disorder}
%%%%%%%%%%%%%%%%%%%%%%%%%%%%%%%%%%%%%%%%%%%%%%%%%%%%%%%%%
%%%%%%%%%%%%%%%%%%%%%%%%%%%%%%%%%%%%%%%%%%%%%%%%%%%%%%%%%
%%%%%%%%%%%%%%%%%%%%%%%%%%%%%%%%%%%%%%%%%%%%%%%%%%%%%%%%
For general values of the frequency the Liapunov exponent $\gamma(\omega)$ is not available 
in explicit analytical form. However, Matsuda et al. \cite{Matsuda68,Matsuda70}
have determined $\gamma(\omega)$ in the low frequency limit in the case of mass disorder. They find
\begin{eqnarray}
\gamma(\omega)\simeq\frac{1}{8}\frac{\omega^2}{\kappa\langle m\rangle}\langle\delta m^2\rangle,
\label{liap4}
\end{eqnarray}
where the mean mass and the mean square mass deviations are given by  $\langle m\rangle$ and 
$\langle\delta m^2\rangle=\langle(m-\langle m\rangle)^2\rangle$; the averages $\langle\cdots\rangle$
determined by the mass distribution $\pi_1(m_n)$.

Here we consider the evaluation of the Liapunov exponent in the case of coupling strength disorder.
Rather than attempting to apply the techniques by Matsuda et al. we here use an approach
advanced by Lepri et al.  \cite{Lepri03} using dynamical system theory  \cite{Jackson90} and 
statistical physics \cite{Reichl98}. From a theoretical physics point a view we believe this method 
is simpler and more straightforward.
%%%%%%%%%%%%%%%%%%%%%%%%%%%%%%%%%%%%%%%%%%%%%%%%%%%%%%%%%
%%%%%%%%%%%%%%%%%%%%%%%%%%%%%%%%%%%%%%%%%%%%%%%%%%%%%%%%%
\subsection{The case: $\omega=0$, $m_n=m$,  $\kappa_n=\kappa$}
\begin{figure}
\includegraphics[width=1.0\hsize]{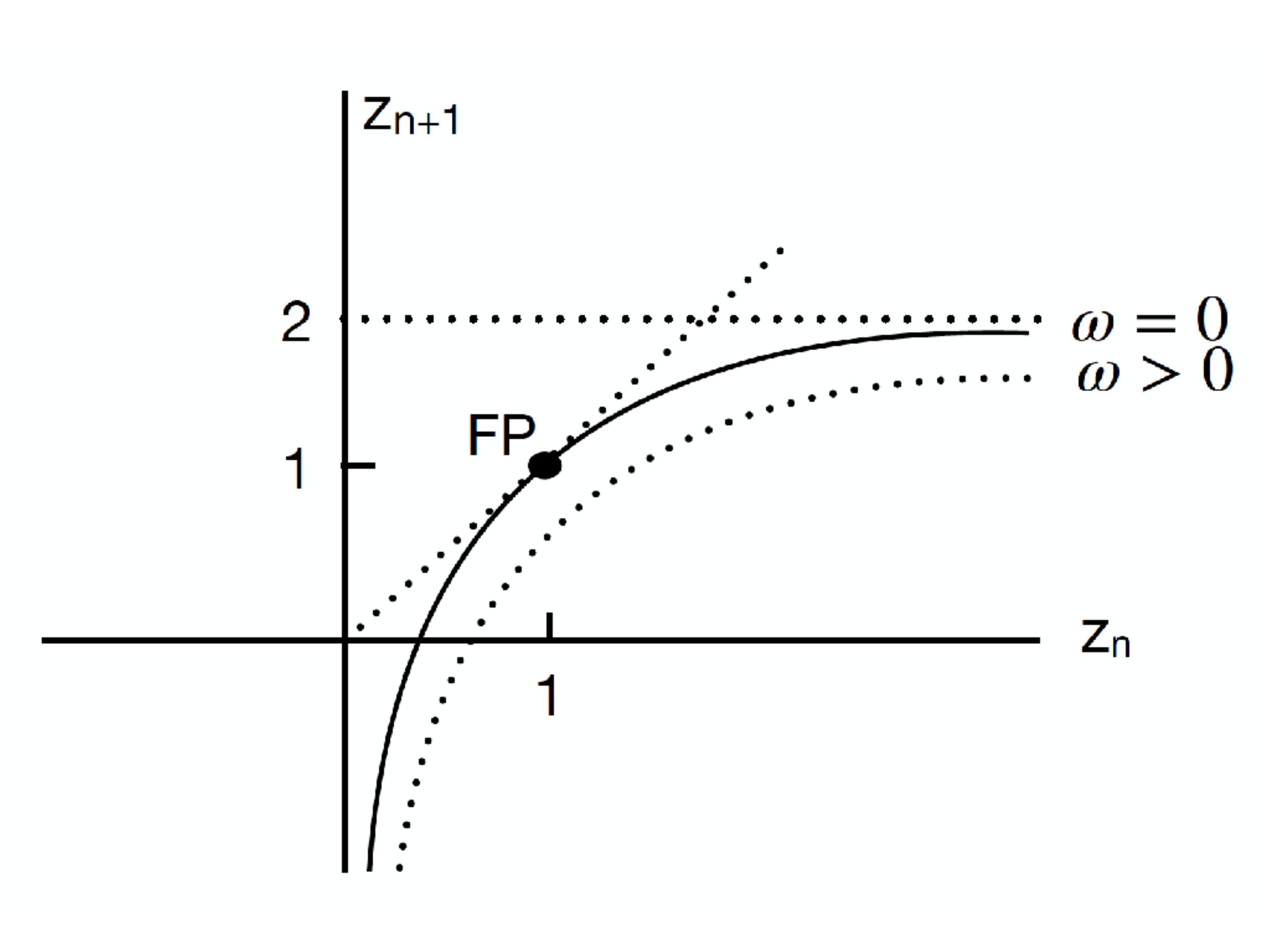}
\caption{Depiction of the phase diagram of the discrete nonlinear map in a plot of $z_{n+1}$ versus
$z_n$. The map has a marginally stable fixed point (FP) at $z\ast=1$. The full curve for $\omega=0$ 
corresponds to the map (\ref{map2}); the dotted curve for $\omega>0$ corresponds to the map
(\ref{map3}).} 
\label{fig2}
\end{figure}

\begin{figure}
\includegraphics[width=1.0\hsize]{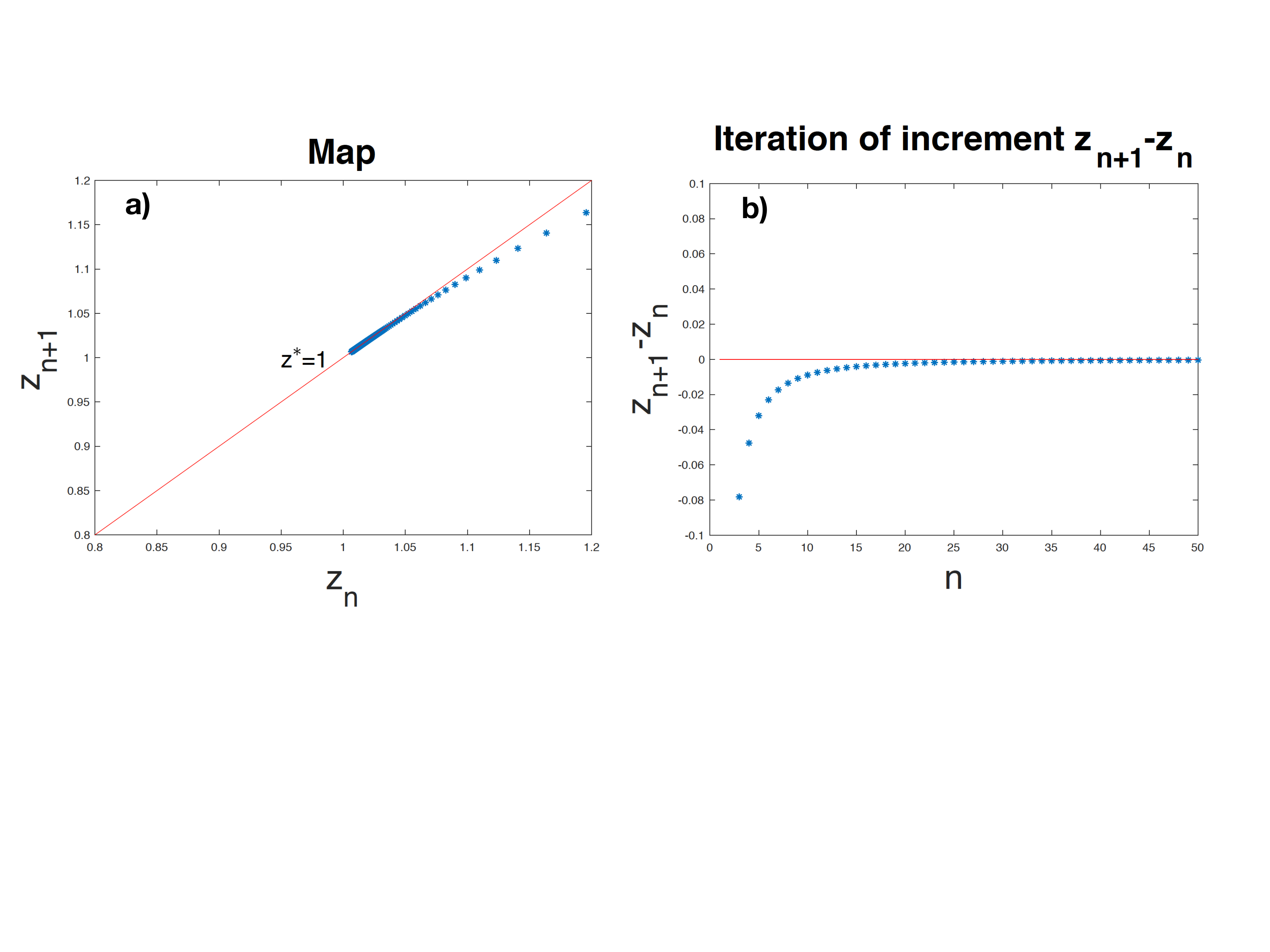}
\caption{In a) we plot $z_{n+1}$ versus $z_n$ and depict the iteration of the map for 
$\omega=0$ in (\ref{map2}). The iterates converge towards the marginally stable fixed point (FP) 
at $z\ast=1$. In b) we depict the increments as function of $n$, showing the convergence to the
fixed point.}
\label{fig3}
\end{figure}

%%%%%%%%%%%%%%%%%%%%%%%%%%%%%%%%%%%%%%%%%%%%%%%%%%%%%%%%%
%%%%%%%%%%%%%%%%%%%%%%%%%%%%%%%%%%%%%%%%%%%%%%%%%%%%%%%%%
For $\omega=0$ and vanishing disorder, i.e., $\kappa_n=\kappa$ and $m_n=m$ the map (\ref{map1}) 
takes the form
\begin{eqnarray}
z_{n+1}=f(z_n)=2-\frac{1}{z_n},
\label{map2}
\end{eqnarray}
where we have omitted the $\omega$ dependence. In a plot of $z_{n+1}$ versus $z_n$ the map
is composed of two hyperbolic branches. For $|z_n|\rightarrow\infty$ we have $z_{n+1}\rightarrow 2$, for 
$z_n\rightarrow\pm 0$ we note that $z_{n+1}\rightarrow\mp\infty$. The map has a fixed point $z^\ast$  determined 
by $f(z^\ast)=z^\ast$, yielding $z^\ast=1$. The evolution of the iterates as a function of $n$ is analysed  
by considering the increment $z_{n+1}-z_n=-(z_n-1)^2/z_n$. We find that $z_{n+1}-z_n<0$ for $z_n>0$, whereas for
$z_n<0$ the increment $z_{n+1}-z_n>0$. At the fixed point $z^\ast=1$ the increment
vanishes, i.e, $z_{n+1}-z_n=0$. In other words, as we approach the fixed point through
iterates $z_n>1$ the iterates converge to the fixed point; on the other hand, choosing
an initial iterate $z_n\lesssim 1$ the iterates move away from the fixed point, corresponding
to a marginally stable fixed point. Further inspection of the map in a plot of $z_{n+1}$
versus $z_n$ shows that choosing an initial value $z_n<1$ the iterates eventually make a 
single excursion to the hyperbola $2-1/z_n$ for $z_n<0$ before returning to the hyperbola
for $z_n>0$ and approaching the fixed point.  In Fig.~\ref{fig2} we have plotted $z_{n+1}$ versus $z_n$ with the
fixed point indicated at $z_{n+1}=z_n=1$; the solid line depicts the map (\ref{map2}).
In Fig.~\ref{fig3} we have in a)  plotted $z_{n+1}$ versus $z_n$ demonstrating the convergence towards 
the fixed point at $z^\ast=1$; in b) we have plotted the decreasing increments $z_{n+1}-z_n$ versus $n$.
%%%%%%%%%%%%%%%%%%%%%%%%%%%%%%%%%%%%%%%%%%%%%%%%%%%%%%%%%
%%%%%%%%%%%%%%%%%%%%%%%%%%%%%%%%%%%%%%%%%%%%%%%%%%%%%%%%%
\subsection{The case: $\omega\gtrsim 0$, $m_n=m$,  $\kappa_n=\kappa$ }
\begin{figure}
\includegraphics[width=1.0\hsize]{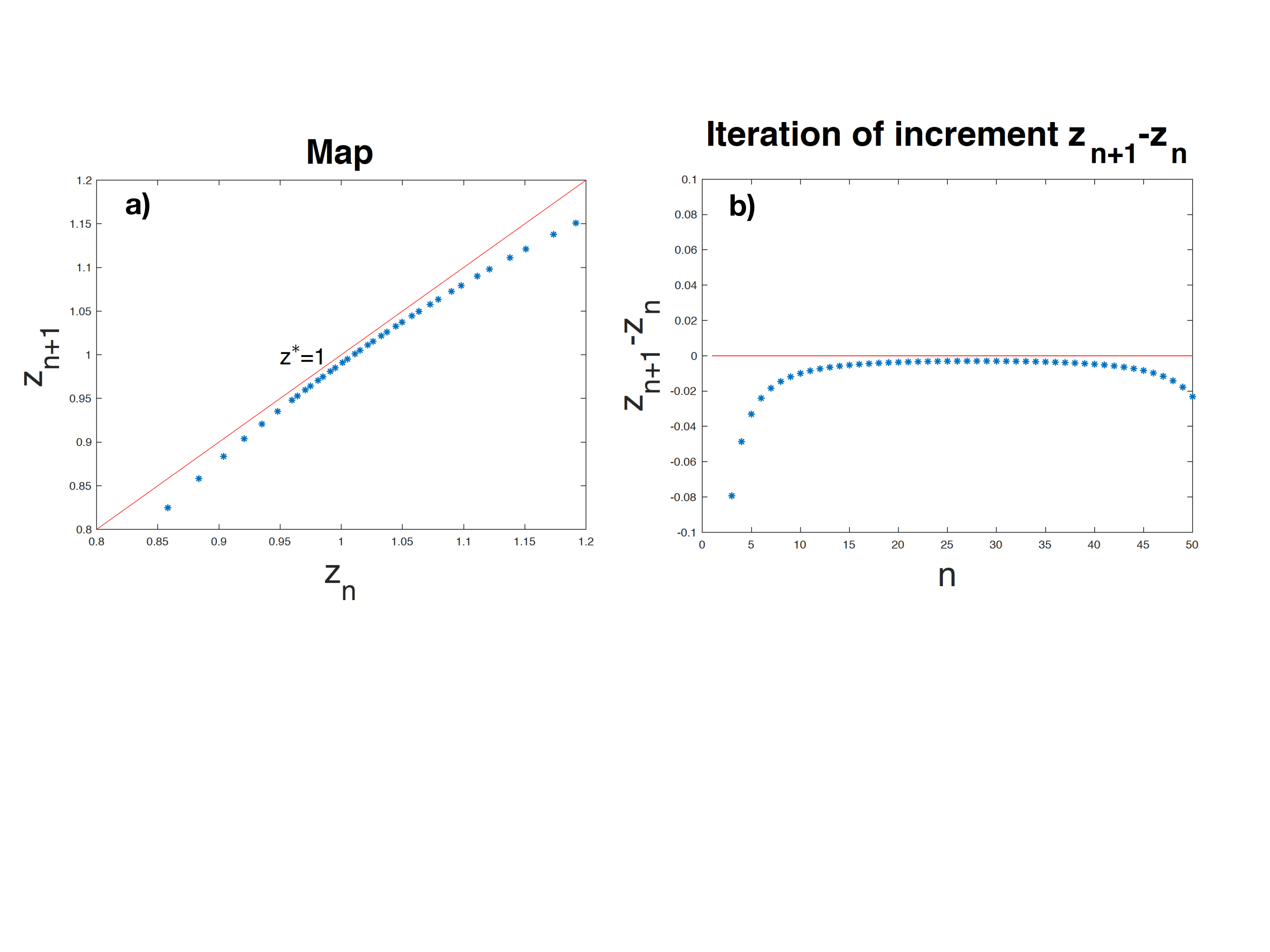}
\caption{In a) we plot $z_{n+1}$ versus $z_n$ and depict the iteration of the map for 
$\omega>0$ in (\ref{map3}). The iterates compress and flow past the point $z=1$ 
(the position of fixed point (FP) $z\ast=1$ for $\omega=0$). In b) we depict the increments as function of $n$, showing the compression near the the point $z=1$, allowing for a continuum approximation.}
\label{fig4}
\end{figure}

%%%%%%%%%%%%%%%%%%%%%%%%%%%%%%%%%%%%%%%%%%%%%%%%%%%%%%%%%
%%%%%%%%%%%%%%%%%%%%%%%%%%%%%%%%%%%%%%%%%%%%%%%%%%%%%%%%%
We next consider the case of small $\omega$ and vanishing disorder. From (\ref{map1}) we infer the map
\begin{eqnarray}
z_{n+1}=2-\frac{m}{\kappa}\omega^2-\frac{1}{z_n},
\label{map3}
\end{eqnarray}
depicted by a dotted line in Fig.~\ref{fig2}. In this case the map does not have a (real) fixed point. However, analysing the increment $z_{n+1}-z_n=2-(m/\kappa)\omega^2-z_n-1/z_n$
in the vicinity of the value $z=1$ (the position of the fixed point for $\omega=0$) we obtain $z_{n+1}-z_n\approx -(m/\kappa)\omega^2$ and the increment 
vanishes for small $\omega$. We note that the increment is negative corresponding to a flux of iterates close to the 
point $z=1$ from the region $z_n>1$ to $z_n<1$. In Fig.~\ref{fig4} we have in a)  plotted $z_{n+1}$ versus $z_n$ demonstrating the 
flux of iterates past the point  $z=1$; in b) we have shown that the increments $z_{n+1}-z_n$ as function
of $n$ decrease in the vicinity of the point $z=1$.
%%%%%%%%%%%%%%%%%%%%%%%%%%%%%%%%%%%%%%%%%%%%%%%%%%%%%%%%%
%%%%%%%%%%%%%%%%%%%%%%%%%%%%%%%%%%%%%%%%%%%%%%%%%%%%%%%%%
\subsection{The case: $\omega\gtrsim 0$, $m_n=m$,  $\delta\kappa_n\approx0$}
%%%%%%%%%%%%%%%%%%%%%%%%%%%%%%%%%%%%%%%%%%%%%%%%%%%%%%%%%
%%%%%%%%%%%%%%%%%%%%%%%%%%%%%%%%%%%%%%%%%%%%%%%%%%%%%%%%%
Finally, we consider the case of small $\omega$ and small coupling strength disorder. Setting 
$\kappa_n=\langle\kappa\rangle+\delta\kappa_n$, where $\langle\kappa\rangle$ is determined by the coupling strength distribution $\pi_2(\kappa_n)$,
and assuming $\delta\kappa_n\ll\langle\kappa\rangle$, we obtain to leading order expanding the map (\ref{map1}) 
\begin{eqnarray}
z_{n+1}=
2-\frac{m}{\langle\kappa\rangle}\bigg(1-\frac{\delta\kappa_n}{\langle\kappa\rangle}\bigg)\omega^2
+\frac{\delta\kappa_{n-1}-\delta\kappa_n}{\langle\kappa\rangle}-
\bigg(1+\frac{\delta\kappa_{n-1}-\delta\kappa_n}{\langle\kappa\rangle}\bigg)\frac{1}{z_n}.
\label{map4}
\end{eqnarray}
Furthermore,  expanding about the point $z=1$ by setting $z_n=1+\epsilon_n$ we note
that the terms $(\delta\kappa_{n-1}-\delta\kappa_n)/\langle\kappa\rangle$ cancels out and we obtain
for small $\epsilon_n$
\begin{eqnarray}
\epsilon_{n+1}-\epsilon_n\simeq -\epsilon_n^2-\frac{m}{\langle\kappa\rangle}\omega^2
+\frac{m\omega^2}{\langle\kappa\rangle^2}\delta\kappa_n.
\label{map5}
\end{eqnarray}
For large $n$  the iterates compress and constitute a flow near the point $z=1$ 
in the sense that $\epsilon_n-\epsilon_{n+1}\rightarrow 0$ for small $\omega$. 
As a consequence we can introduce the continuum limit and make the assumption $\epsilon_n\approx\epsilon(n)$ 
and $\delta\kappa_n\approx\delta\kappa(n)$, where $n$ is a continuous variable. From (\ref{map5}) we thus obtain 
the effective Langevin equation
\begin{eqnarray}
\frac{d\epsilon(n)}{dn}&=&-\epsilon(n)^2-\frac{m}{\langle\kappa\rangle}\omega^2+\eta(n)
\label{lan3},
\\
\eta(n)&=&\frac{m\omega^2}{\langle\kappa\rangle^2}\delta\kappa(n),
\label{noi3}
\end{eqnarray}
where we have introduced the 'noise variable' $\eta(n)$ with correlations
\begin{eqnarray}
&&\langle\eta(n)\eta(n')\rangle=\Delta\delta(n-n'),
\label{corkap}
\\
&&\Delta=\bigg(\frac{m\omega^2}{\langle\kappa\rangle^2}\bigg)^2\langle\delta\kappa^2\rangle.
\end{eqnarray}
Expressing the Langevin equation in (\ref{lan3}) in the form $d\epsilon/dn=-dV/d\epsilon+\eta$ the 'potential'
has the form $V=(m/\langle\kappa\rangle)\omega^2\epsilon+\epsilon^3/3$ with a linear slope
for small $\epsilon$. Since there is no minimum the 'position' $\epsilon$ 'falls down' the slope and
escapes for negative $\epsilon$. This is
consistent with the behaviour of the iterates near the the point $z=1$  where there is a flow from right to left
implying that the stochastic map generates a probability current $J_0$ near $z=1$.

In order to proceed we assume that the coupling strength distribution $\pi_2(\kappa)$ has a Gaussian form, implying
that the 'noise' driving the Langevin equation (\ref{lan3}) has the structure of Gaussian white noise \cite{Reichl98}.
This implies that the probability density $P(\epsilon,n)$ is governed by the Fokker-Planck equation
\cite{Risken89} 
\begin{eqnarray}
\frac{\partial P(\epsilon,n)}{\partial n}=
\frac{\partial}{\partial\epsilon}\Big(\epsilon^2+\frac{m}{\langle\kappa\rangle}\omega^2\Big)P(\epsilon,n)
+\frac{1}{2}\Delta\frac{\partial^2P(\epsilon,n)}{\partial\epsilon^2}.
\label{fp}
\end{eqnarray}
From the continuity equation $\partial P/\partial n=-\partial J_0/\partial\epsilon$ we identify
the probability current 
\begin{eqnarray}
J_0=-\Big(\epsilon^2+\frac{m}{\langle\kappa\rangle}\omega^2\Big)P(\epsilon,n)
-\frac{1}{2}\Delta\frac{\partial P(\epsilon,n)}{\partial\epsilon}.
\label{pc}
\end{eqnarray}
For small $\omega$ the stationary distribution to leading asymptotic order in $\Delta$ has the form
\begin{eqnarray}
P_0(\epsilon)\propto\frac{1}{\epsilon^2+(m/\langle\kappa\rangle)\omega^2}+
\Delta\frac{\epsilon}{(\epsilon^2+(m/\langle\kappa\rangle)\omega^2)^3},
\label{stat}
\end{eqnarray}
for a technical detail see Appendix \ref{liapunovapp}.
Finally, from (\ref{liap2}), expanding $\log(|z(\omega)|)\sim 1+\epsilon(\omega)$, we have
\begin{eqnarray}
\gamma(\omega)=\frac{\int d\epsilon \epsilon P_0(\epsilon)}{\int d\epsilon P_0(\epsilon)}.
\label{liap5}
\end{eqnarray}
Inserting $P_0(\epsilon)$ we note that the first term in (\ref{stat}) for symmetry reasons does not contribute and 
we obtain by quadrature to order $\Delta$ the Liapunov exponent for small $\omega$ for a coupling strength-disordered chain
\begin{eqnarray}
\gamma(\omega)\simeq
\frac{1}{8}\frac{\omega^2}{\langle\kappa\rangle m}
\bigg(\frac{m}{\langle\kappa\rangle}\bigg)^2\langle\delta\kappa^2\rangle.
\label{liap6}
\end{eqnarray}
%
%%%%%%%%%%%%%%%%%%%%%%%%%%%%%%%%%%%%%%%%%%%%%%%%%%%%%%%%%
%%%%%%%%%%%%%%%%%%%%%%%%%%%%%%%%%%%%%%%%%%%%%%%%%%%%%%%%%
\subsection{Combining coupling strength disorder and mass disorder}
%%%%%%%%%%%%%%%%%%%%%%%%%%%%%%%%%%%%%%%%%%%%%%%%%%%%%%%%%
%%%%%%%%%%%%%%%%%%%%%%%%%%%%%%%%%%%%%%%%%%%%%%%%%%%%%%%%%
Here we consider as a corollary the case of both coupling strength disorder and mass disorder. 
In the map (\ref{map1}) we note that in the absence of coupling strength disorder the random mass 
$m_n$ multiplies $\omega$ and is quenched in the low frequency limit. In the analysis by Matsuda et al.
 \cite{Matsuda68,Matsuda70} the Liapunov exponent in (\ref{liap4}) only depends on the first and second 
moment of the mass distribution, i.e., $\langle m\rangle$ and $\langle\delta m^2\rangle$.
In other words, the calculation of (\ref{liap4}) does not presuppose a narrow mass distribution.

Including mass disorder in the Langevin equation in (\ref{lan3}) by setting $m(n)=\langle m\rangle+\delta m(n)$
we obtain
\begin{eqnarray}
\frac{d\epsilon(n)}{dn}&=&-\epsilon(n)^2-\frac{\langle m\rangle}{\langle\kappa\rangle}\omega^2+\tilde\eta(n)
\label{lan4},
\\
\tilde\eta(n)&=&\frac{\langle m\rangle\omega^2}{\langle\kappa\rangle^2}
\delta\kappa(n)-\frac{\omega^2}{\langle\kappa\rangle}\delta m(n)
\label{noi4}.
\end{eqnarray}
Ignoring terms of order $\delta m\delta\kappa$ we obtain the noise correlations
\begin{eqnarray}
&&\langle\tilde\eta(n)\tilde\eta(n')\rangle=\tilde\Delta\delta(n-n'),
\label{corkapr}
\\
&&\tilde\Delta=\bigg(\frac{\omega^2}{\langle\kappa\rangle}\bigg)^2
\bigg[\langle\delta m^2\rangle+\bigg(\frac{\langle m\rangle}{\langle\kappa\rangle}\bigg)^2\langle\delta\kappa^2\rangle\bigg],
\label{newdel}
\end{eqnarray}
and correspondingly the Liapunov exponent
\begin{eqnarray}
&&\tilde\gamma(\omega)\simeq
\frac{1}{8}\frac{\omega^2}{\langle\kappa\rangle\langle m\rangle}\langle\delta\tilde m^2\rangle,
\label{liap7}
\\
&&\langle\delta\tilde m^2\rangle=
\langle\delta m^2\rangle+\bigg(\frac{\langle m\rangle}{\langle\kappa\rangle}\bigg)^2\langle\delta\kappa^2\rangle.
\label{r-rmqn}
\end{eqnarray}
For $\delta\kappa =0$ we recover the Liapunov exponent in the mass-disordered case in (\ref{liap4})
first derived by Matsuda et al. \cite{Matsuda68,Matsuda70}, see also \cite{Lepri03}. The presence of weak coupling strength disorder
can be incororated by introducing the renormalised mean square mass deviation  $\langle\delta\tilde m^2\rangle$
given by (\ref{r-rmqn}). The expressions in (\ref{liap6}) and (\ref{liap7}-\ref{r-rmqn}) constitute the main results of 
the present analysis. 
%%%%%%%%%%%%%%%%%%%%%%%%%%%%%%%%%%%%%%%%%%%%%%%%%%%%%%%%%
%%%%%%%%%%%%%%%%%%%%%%%%%%%%%%%%%%%%%%%%%%%%%%%%%%%%%%%%%
\subsection{Connection to 1D quantum mechanical disorder}
%%%%%%%%%%%%%%%%%%%%%%%%%%%%%%%%%%%%%%%%%%%%%%%%%%%%%%%%%
%%%%%%%%%%%%%%%%%%%%%%%%%%%%%%%%%%%%%%%%%%%%%%%%%%%%%%%%%
The Langevin equations in (\ref{lan3}) and (\ref{lan4}) have the form of a Ricatti equation of the form
$\epsilon'=-\epsilon^2-E+V$, where $V=\tilde\eta$ and $E=\omega^2\langle m\rangle/\langle\kappa\rangle$,
the prime denoting a derivative. By means of the substitution $\epsilon=\psi'/\psi$ the Ricatti equation is 
reduced to the 1D stationary Schr\"{o}dinger equation, $-\psi''+V\psi=E\psi$, describing the quantum motion in a random
potential $V$ with zero mean and "white noise"correlations 
$\langle\tilde\eta(n)\tilde\eta(n')\rangle=\tilde\Delta\delta(n-n')$. This problem, relating to Anderson localisation
\cite{Anderson58}, has been studied extensively, see e.g.  \cite{Nieuwenhuizen83,Lifshits88,Luck04,Grabsch14}.
In 1D in the presence of even weak disorder the wave function is localised, characterised by the localisation or
correlation length $l_c$. The corresponding Liapunov exponent is thus given by $\gamma=1/l_c$. According to
the analysis by Luck \cite{Luck04} one finds in the case of weak disorder $\gamma=\tilde\Delta/8E$ and by insertion
the result in (\ref{liap7}).
%%%%%%%%%%%%%%%%%%%%%%%%%%%%%%%%%%%%%%%%%%%%%%%%%%%%%%%%%
%%%%%%%%%%%%%%%%%%%%%%%%%%%%%%%%%%%%%%%%%%%%%%%%%%%%%%%%%
%%%%%%%%%%%%%%%%%%%%%%%%%%%%%%%%%%%%%%%%%%%%%%%%%%%%%%%%%
\section{\label{heatldf}Heat Current and heat fluctuations}
%%%%%%%%%%%%%%%%%%%%%%%%%%%%%%%%%%%%%%%%%%%%%%%%%%%%%%%%%
%%%%%%%%%%%%%%%%%%%%%%%%%%%%%%%%%%%%%%%%%%%%%%%%%%%%%%%%%
%%%%%%%%%%%%%%%%%%%%%%%%%%%%%%%%%%%%%%%%%%%%%%%%%%%%%%%%%
Here we briefly discuss the implication of a Liapunov exponent for the heat current and heat fluctuations.
%%%%%%%%%%%%%%%%%%%%%%%%%%%%%%%%%%%%%%%%%%%%%%%%%%%%%%%%%
%%%%%%%%%%%%%%%%%%%%%%%%%%%%%%%%%%%%%%%%%%%%%%%%%%%%%%%%%
%%%%%%%%%%%%%%%%%%%%%%%%%%%%%%%%%%%%%%%%%%%%%%%%%%%%%%%%%
\subsection{Heat Current}
%%%%%%%%%%%%%%%%%%%%%%%%%%%%%%%%%%%%%%%%%%%%%%%%%%%%%%%%%
%%%%%%%%%%%%%%%%%%%%%%%%%%%%%%%%%%%%%%%%%%%%%%%%%%%%%%%%%
%%%%%%%%%%%%%%%%%%%%%%%%%%%%%%%%%%%%%%%%%%%%%%%%%%%%%%%%%
Regarding the heat current we summarise the analysis by Dhar \cite{Dhar01,Dhar08a,Dhar16} below.
According to (\ref{qr1}) the fluctuating heat rate from reservoir 1 is given by 
$\dot Q(t)=(-\Gamma\dot u_1(t)+\xi_1(t))\dot u_1(t)$ and the integrated heat flux by $Q(t)=\int^tdt'\dot Q(t')$.
Averaged over the heat reservoirs the mean value $\langle Q(t)\rangle\propto t$ and we obtain the mean
heat current $J(N)=\langle Q(t)\rangle/t$, $\langle\cdots\rangle$ denoting a thermal average.

For an ordered chain the mean heat current is given by the expression 
\cite{Casher71,Rubin71,Dhar06,Dhar08a,Fogedby12}
\begin{eqnarray}
J(N)=\frac{1}{2}(T_1-T_2)\int\frac{d\omega}{2\pi}T(\omega),
\label{heat1}
\end{eqnarray}
where the transmission matrix $T(\omega)$ is expressed in terms of the end-to-end Green's function in (\ref{green1}),
\begin{eqnarray}
T(\omega)=4(\omega\Gamma)^2|G_{1N}(\omega)|^2.
\label{trans2}
\end{eqnarray}
Since $T(\omega)$ is bounded and the range of $\omega$ is determined by the phonon dispersion law  (\ref{disp})
it follows that the heat current $J\propto (T_1-T_2)$, yielding a conductivity $\kappa\propto N$; this behaviour
is characteristic of ballistic heat transport. 

In Appendix \ref{current} we have derived the expression (\ref{heat1}), see also \cite{Fogedby12}, 
and find  that it also holds for the
disordered chain with the Green's function (\ref{green2}) for a particular disorder realisation 
$\{m_n\}$ and $\{\kappa_n\}$. The issue of averaging the current $J(N)$ with respect to the disorder
is, however, quite complex and we review the analysis by Dhar \cite{Dhar01,Dhar08a,Dhar16} here.

For a system of size $N$ the matrix elements $B_{ij}(\omega)$ in $G_{1N}(\omega)$ scale according
to (\ref{bom2}) like $\exp(\tilde\gamma(\omega)N)$, where $\tilde\gamma(\omega)$ 
for small $\omega$ and both mass disorder and weak coupling strength disorder is given by (\ref{liap7})
and (\ref{r-rmqn}) . 
Consequently, for  $\tilde\gamma(\omega) N\gg 1$ the denominator in $G_{1N}(\omega)$ diverges and the 
heat current vanishes; this is due to the localised modes which do not carry energy and thus do not contribute 
to the heat transport. On the other hand, for $\tilde\gamma(\omega) N\ll 1$, corresponding to the extended modes, 
the Green's function $G_{1N}(\omega)$ is bounded and contributes to the heat current. 

The limiting case for $\tilde\gamma(\omega) N\approx 1$ defines the correlation or cross-over frequency
\begin{eqnarray}
\omega_c=(8\langle\kappa\rangle\langle m\rangle)^{1/2}(\langle\delta\tilde m^2\rangle N)^{-1/2}.
\label{cross}
\end{eqnarray}
Consequently, the integration over frequencies in (\ref{heat1}) is cut-off at $\omega=\omega_c$.
An approximate expression for the disorder-averaged heat current, characterised by a bar, is thus
given by 
\begin{eqnarray}
\overline{J(N)}\simeq\frac{1}{2}(T_1-T_2)\int_{-\omega_c}^{\omega_c}\frac{d\omega}{2\pi}T(\omega).
\label{heat2}
\end{eqnarray}
Owing to the $N$-dependence of the cut-off frequency $\omega_c$ the heat current  $\overline{J(N)}$ acquires an explicit $N$ dependence. In the range $|\omega|<\omega_c$ of extended modes Dhar \cite{Dhar01,Dhar08a} uses for $G_{1N}(\omega)$ the unperturbed result given by (\ref{green1}). This is an excellent approximation supported by numerical estimates \cite{Dhar08a}.  
Since $T(\omega)\sim\omega^2$ for small $\omega$ a simple scaling argument yields $\overline{J(N)}\sim \omega_c^3\sim(\langle\delta\tilde m^2\rangle N)^{-3/2} $, corresponding to the exponent $\alpha=3/2$; 
note that in the ballistic case $\alpha=0$. We also find  that the heat current  for large fixed $N$ scales with the mean square
renormalised mass according to $\overline{J(N)}\sim\langle\delta\tilde m^2\rangle^{-3/2}$.
%%%%%%%%%%%%%%%%%%%%%%%%%%%%%%%%%%%%%%%%%%%%%%%%%%%%%%%%%
%%%%%%%%%%%%%%%%%%%%%%%%%%%%%%%%%%%%%%%%%%%%%%%%%%%%%%%%%
%%%%%%%%%%%%%%%%%%%%%%%%%%%%%%%%%%%%%%%%%%%%%%%%%%%%%%%%%
\subsection{\label{cumulant} Large deviation function}
%%%%%%%%%%%%%%%%%%%%%%%%%%%%%%%%%%%%%%%%%%%%%%%%%%%%%%%%%
%%%%%%%%%%%%%%%%%%%%%%%%%%%%%%%%%%%%%%%%%%%%%%%%%%%%%%%%%
%%%%%%%%%%%%%%%%%%%%%%%%%%%%%%%%%%%%%%%%%%%%%%%%%%%%%%%%%
The distribution of heat fluctuations is described by the moment generating characteristic function \cite{Reichl98}
\begin{eqnarray}
C(\lambda,t)=\langle\exp(\lambda Q(t))\rangle.
\label{char1}
\end{eqnarray}
Correspondingly, the cumulant generating function is given by $\log C(\lambda,t)$. The long 
time behaviour is characterised by the associated large deviation function $\mu(\lambda)$ 
according to
\begin{eqnarray}
C(\lambda,t)=\exp(\mu(\lambda) t).
\label{char2}
\end{eqnarray}
It follows from general principles \cite{Touchette09,Hollander00,Fogedby11a,Fogedby12,Fogedby14b} that the cumulant generating function $\mu(\lambda)$ is downward convex and owing to normalisation passes through the origin, i.e.,  $\mu(0)=0$.

For an ordered chain the large deviation function has been derived by Saito and Dhar \cite{Saito07,Saito11},
see also \cite{Kundu11}. Here we present a derivation in Appendix \ref{cgf}, see also \cite{Fogedby12}.
The large deviation function has the form
\begin{eqnarray}
&&\mu(\lambda)=-\frac{1}{2}\int\frac{d\omega}{2\pi}\log[1+T(\omega)f(\lambda)],
\label{ldf}
\\
&&T(\omega)=4(\omega\Gamma)^2|G_{1N}(\omega)|^2,
\\
&&f(\lambda)=T_1T_2\lambda(1/T_1-1/T_2-\lambda).
\label{f}
\end{eqnarray}
Here the structure of $f(\lambda)$ ensures that $\mu(\lambda)$ satisfies the Gallavotti-Cohen fluctuation theorem \cite{Gallavotti95,Lebowitz99,Fogedby12} valid for driven non equilibrium systems,
\begin{eqnarray}
\mu(\lambda)=\mu(1/T_1-1/T_2-\lambda).
\label{ft}
\end{eqnarray}
From the structure of (\ref{ldf}) it follows that $\mu(\lambda)$ has branch points determined by the condition
$1+T(\omega)f(\lambda)>0$. Since by inspection $0\le T(\omega)\le 1$, see \cite{Fogedby12}, it follows that
$f(\lambda)>-1$ yielding the branch points $\lambda_1= 1/T_1$ and $\lambda_2= -1/T_2$. We also note that
the fluctuation theorem in (\ref{ft}) implies that $\mu(1/T_1-1/T_2)=0$. In conclusion, the large deviation function
is downward convex, crossing the axis at $\lambda=0$ and $\lambda =1/T_1-1/T_2$ and having branch points at
$\lambda_1$ and $\lambda_2$. At equal temperature $T_1=T_2$ the large deviation function is positive in the
whole range as shown in Fig.~\ref{fig6}.

The expression for $\mu(\lambda)$ in (\ref{ldf}) is for a concrete realisation of the quenched mass and coupling 
strength disorder $\{m_n\}$ and $\{\kappa_n\}$  through the dependence on the Green's function 
$G_{1N}(\omega)$ in (\ref{green2}). We note, however, that due to the form of $f(\lambda)$ the large deviation function 
$\mu(\lambda)$ satisfies the Gallavotti-Cohen fluctuation theorem for each disorder realisation and we conclude
that the disorder-averaged large deviation function $\overline{\mu(\lambda)}$, likewise, obeys the fluctuation
theorem.

A further clarification also follows from the Fokker-Planck equation for the joint
distribution $P(Q_1,Q_2, \{u_n\}, \{\dot u_n\},t)$ for the heat transfer \cite{Imparato07,Fogedby12,Risken89}
\begin{eqnarray}
\frac{\partial P}{\partial t} = &&(L_0+L_Q)P,
\label{fp1}
\\
L_0 P=&&\{P,H\},
\label{fp2}
\\
L_QP=&&\Gamma\Bigg(T_1\dot u_1^2\frac{\partial^2P}{\partial Q_1^2}+
2T_1\dot u_1\frac{\partial^2P}{\partial Q_1\partial\dot u_1}+(\dot u_1^2+T_1)\frac{\partial P}{\partial Q_1}\Bigg)
\nonumber
\\
+&&\Gamma\Bigg(T_2\dot u_N^2\frac{\partial^2P}{\partial Q_2^2}+
2T_2\dot u_N\frac{\partial^2P}{\partial Q_2\partial\dot u_N}+(\dot u_N^2+T_2)\frac{\partial P}{\partial Q_2}\Bigg).
\label{fp3}
\end{eqnarray}
As shown in \cite{Fogedby12} the fluctuation theorem here follows from the structure of the operator $L_Q$ 
and does not depend on the Hamiltonian part $L_0 P$.  Since the disorder only enters in the Hamiltonian $H$
in (\ref{ham})  we again infer the validity of the fluctuation theorem.

\begin{figure}
\includegraphics[width=1.0\hsize]{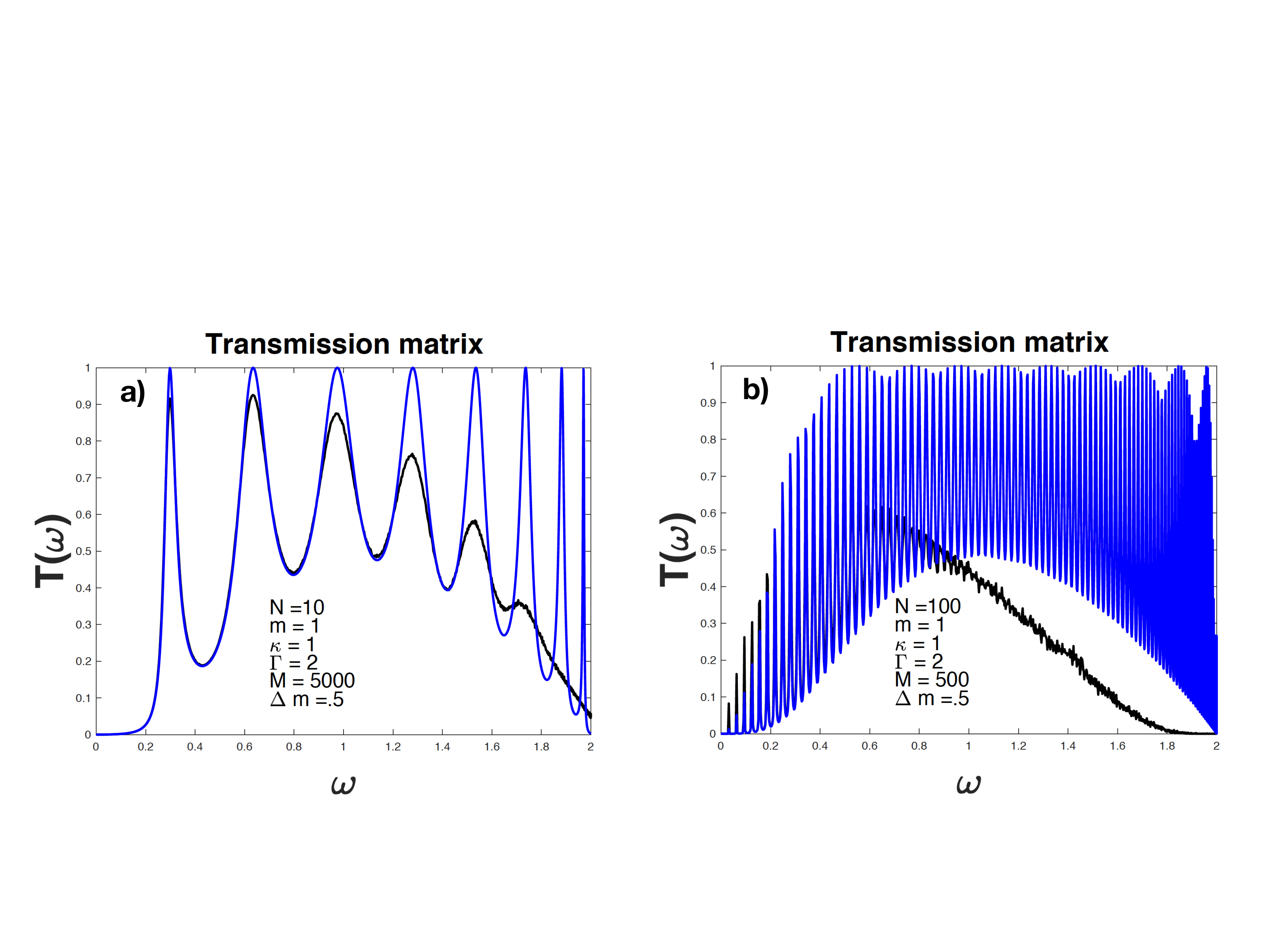}
\caption{In a) we depict the transmission matrix $T(\omega)$ as a function
of $\omega$ for $N=10$, $m=1$, $\kappa=1$, and $\Gamma=2$. The blue curve refers to the 
ordered chain, i.e., in the absence of disorder, showing the resonance structure in $G_{1N}(\omega)$.
The black curve refers to the disordered case for $\Delta m=0.5$ averaged over
$M=5000$ samples. In b)  we depict $T(\omega)$ for $N=100$ and $M=500$. In both cases 
black curve displays the reduction of $T(\omega)$ for larger frequencies due to the disorder.}
\label{fig5}
\end{figure}

\begin{figure}
\includegraphics[width=0.7\hsize]{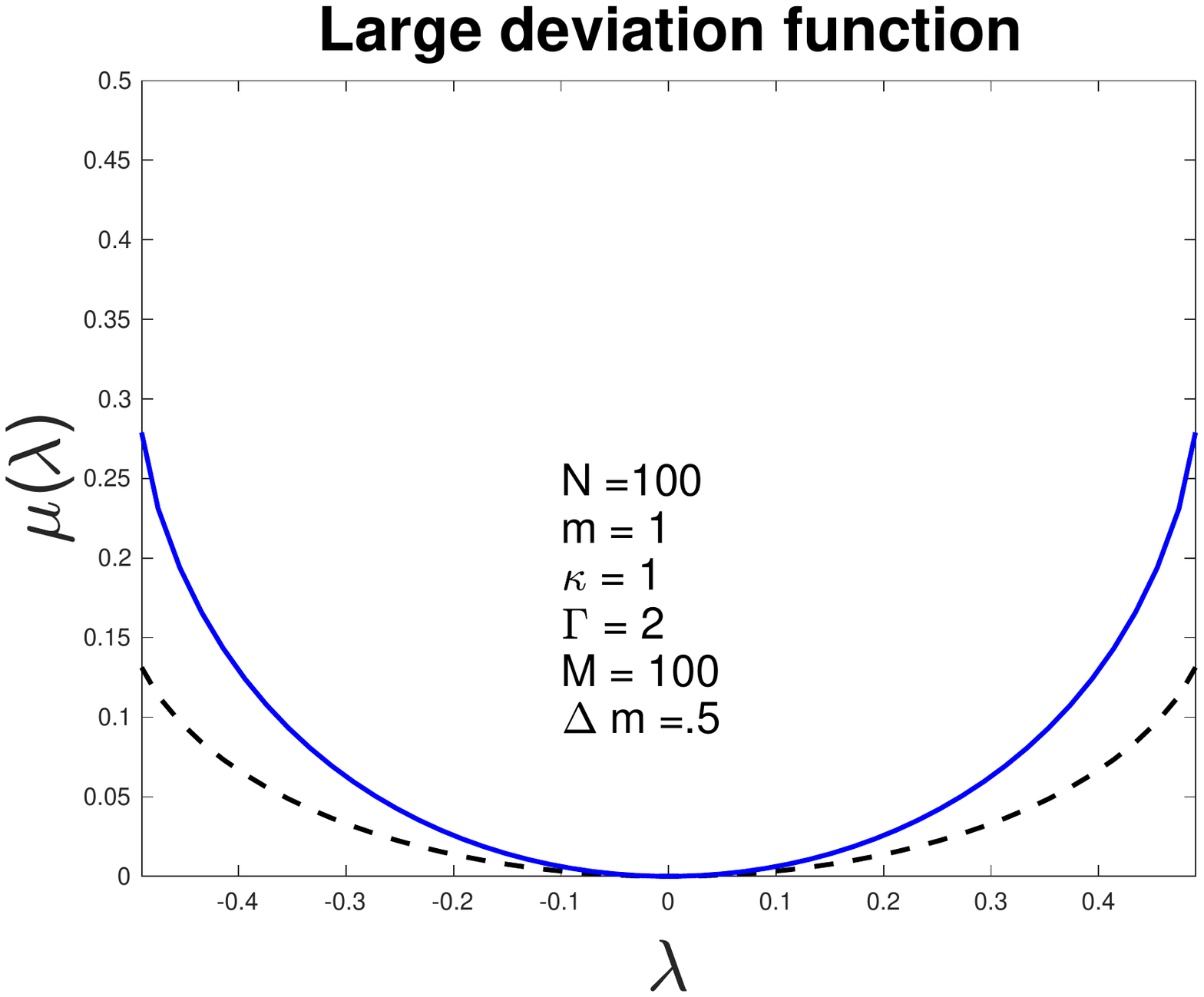}
\caption{We depict the large deviation function $\mu(\lambda)$ as a function of $\lambda$ for $N=100$,
$m=1$, $\kappa=1$, and $\Gamma=2$. The blue curve refers to the ordered chain, i.e., in the
absence of disorder. The black dashed curve refers to the disordered case for $\Delta m=0.5$ averaged over
$M=100$ samples. Since $T(\omega)$ is reduced in the upper $\omega$ range the large deviation function sampling 
all frequencies is overall reduced. However, since $f(\lambda)$ has the form of an inverted parabola, 
the reduction of $\overline{\mu(\lambda)}$ is most pronounced for $\lambda$ close to the edges.}
\label{fig6}
\end{figure}

The disorder enters in the transmission matrix $T(\omega)$ given by (\ref{trans2}). 
In Fig.~\ref{fig5} we have in a) depicted the transmission matrix for $N=10$, $m=1$, $\kappa=1$, and
$\Gamma=2$. The blue curve refers to the ordered case for $\langle\delta\tilde m^2\rangle=0$,
showing the resonance structure of $G_{1N}(\omega)$. The black curve corresponds to the disordered
case for $\sqrt{\langle\delta\tilde m^2\rangle}=0.5$ averaged over 5000 samples. With this choice of parameters
the cross-over frequency in (\ref{cross}) is $\omega_c\sim 1.8$ in accordance with Fig.~\ref{fig5}, showing 
the onset of localised states for $\omega>\omega_c$, yielding a decreasing transmission matrix;
in Fig.~\ref{fig5} we have in b) depicted $T(\omega)$ for $N=100$ and $500$ samples showing the same features.

Finally, implementing the same approximation as for the heat current in (\ref{heat2}), 
we express the disorder-averaged large deviation function in the form
\begin{eqnarray}
\overline{\mu(\lambda)}\simeq-
\frac{1}{2}\int_{-\omega_c}^{\omega_c}\frac{d\omega}{2\pi}\log[1+T(\omega)f(\lambda)],
\label{ldfav}
\end{eqnarray}
where $T(\omega)=4(\omega\Gamma)^2|G_{1N}(\omega)|^2$ is expressed in terms of the
Green's function $G_{1N}(\omega)$ for the ordered case in (\ref{green1}). We have not investigated
the expression in (\ref{ldfav}) further but have determined  $\overline{\mu(\lambda)}$ numerically.
In Fig.~\ref{fig6} we have depicted the large deviation function for $N=100$ both in the absence of disorder
for $\delta\tilde m=0$ and in the presence of disorder choosing $\delta\tilde m=0.5$.  
Since $T(\omega)$ is reduced in the upper $\omega$ range the large deviation function sampling 
all frequencies is overall reduced. However, since $f(\lambda)$ has the form of an inverted parabola, 
the reduction of $\overline{\mu(\lambda)}$ is most pronounced for $\lambda$ close to the edges, 
as shown in Fig.~\ref{fig6}.
%%%%%%%%%%%%%%%%%%%%%%%%%%%%%%%%%%%%%%%%%%%%%%%%%%%%%%%%%
%%%%%%%%%%%%%%%%%%%%%%%%%%%%%%%%%%%%%%%%%%%%%%%%%%%%%%%%%
%%%%%%%%%%%%%%%%%%%%%%%%%%%%%%%%%%%%%%%%%%%%%%%%%%%%%%%%%
\section{\label{conclusion} Conclusion}
%%%%%%%%%%%%%%%%%%%%%%%%%%%%%%%%%%%%%%%%%%%%%%%%%%%%%%%%%
%%%%%%%%%%%%%%%%%%%%%%%%%%%%%%%%%%%%%%%%%%%%%%%%%%%%%%%%%
%%%%%%%%%%%%%%%%%%%%%%%%%%%%%%%%%%%%%%%%%%%%%%%%%%%%%%%%%
In this paper we have discussed the disordered harmonic chain subject to coupling strength disorder.
A case which to our knowledge has not been studied previously. Using a dynamical system theory approach 
we have evaluated the Liapunov exponent at low frequency and for weak coupling strength disorder. 
Including mass disorder we have obtained an expression for the Liapunov exponent which interpolates between 
coupling strength disorder and mass disorder. In the absence of coupling strength disorder we recover the
well-known result by Matsuda et al., see also Lepri. In the general case coupling strength disorder can be
incorporated by introducing a renormalised mass disorder. Finally, we have discussed the heat current and
the large deviation function and commented on the validity of the  Gallavotti-Cohen fluctuation 
theorem for the disordered chain.
%%%%%%%%%%%%%%%%%%%%%%%%%%%%%%%%%%%%%%%%%%%%%%%%%%%%%%%%%
%%%%%%%%%%%%%%%%%%%%%%%%%%%%%%%%%%%%%%%%%%%%%%%%%%%%%%%%%
%%%%%%%%%%%%%%%%%%%%%%%%%%%%%%%%%%%%%%%%%%%%%%%%%%%%%%%%%
\section{\label{appendix}Appendix}
%%%%%%%%%%%%%%%%%%%%%%%%%%%%%%%%%%%%%%%%%%%%%%%%%%%%%%%%%
%%%%%%%%%%%%%%%%%%%%%%%%%%%%%%%%%%%%%%%%%%%%%%%%%%%%%%%%%
%%%%%%%%%%%%%%%%%%%%%%%%%%%%%%%%%%%%%%%%%%%%%%%%%%%%%%%%%
%%%%%%%%%%%%%%%%%%%%%%%%%%%%%%%%%%%%%%%%%%%%%%%%%%%%%%%%%
%%%%%%%%%%%%%%%%%%%%%%%%%%%%%%%%%%%%%%%%%%%%%%%%%%%%%%%%%
%%%%%%%%%%%%%%%%%%%%%%%%%%%%%%%%%%%%%%%%%%%%%%%%%%%%%%%%%
\subsection{\label{greensfunction}Greens function}
%%%%%%%%%%%%%%%%%%%%%%%%%%%%%%%%%%%%%%%%%%%%%%%%%%%%%%%%%
%%%%%%%%%%%%%%%%%%%%%%%%%%%%%%%%%%%%%%%%%%%%%%%%%%%%%%%%%
%%%%%%%%%%%%%%%%%%%%%%%%%%%%%%%%%%%%%%%%%%%%%%%%%%%%%%%%%
Inserting (\ref{om}) we have in Fourier space the Langevin equations
\begin{eqnarray}
&&\Omega_1(\omega)\tilde u_1(\omega)=i\omega\Gamma\tilde  u_1(\omega)+
\kappa_1\tilde u_2(\omega)+\tilde\xi_1(\omega),
\label{a1}
\\
&&\Omega_N(\omega)\tilde u_N(\omega)=i\omega\Gamma\tilde u_N(\omega)+
\kappa_{N-1}\tilde u_{N-1}(\omega)+\tilde \xi_2(\omega),
\label{a2}
\end{eqnarray}
and we infer that the inverse Green's function in (\ref{eqmo2}) has the form 
\begin{eqnarray}
G^{-1}(\omega)=
\left(\begin{array}{ccccc}
\tilde\Omega_1(\omega) &-\kappa_1 &  &  & 
\\
-\kappa_1 & \Omega_2(\omega) &-\kappa_2 &  & 
 \\
  & -\kappa_2 &\cdots  &  &  
 \\
  &  &  & \Omega_{N-1}(\omega) & -\kappa_{N-1} 
 \\
  &  &  &-\kappa_{N-1} & \tilde\Omega_N(\omega)
 \end{array}\right),
\label{a3}
\end{eqnarray}
where $\tilde\Omega_1(\omega)=\Omega_1(\omega)-i\Gamma\omega$ and 
$\tilde\Omega_N(\omega)=\Omega_N(\omega)-i\Gamma\omega$.
We note that $G^{-1}(\omega)$ is a symmetric tridiagonal matrix. For the matrix elements we thus have
 $G_{nm}^{-1}(\omega)=G_{mn}^{-1}(\omega)$ implying $G_{nm}(\omega)=G_{mn}(\omega)$. 
The symmetry and structure also allows us to derive the useful Schwinger identity \cite{Wang45}. 
From (\ref{a3}) we have, the $\ast$ indicting a complex conjugate, 
\begin{eqnarray}
G_{nm}^{-1}(\omega)-G_{nm}^{-1\ast}(\omega)=
((\tilde\Omega_1(\omega)-\tilde\Omega_1^\ast(\omega))\delta_{n1}+
(\tilde\Omega_N(\omega)-\tilde\Omega_N^\ast(\omega))\delta_{nN})\delta_{nm},
\label{a4}
\end{eqnarray}
or
\begin{eqnarray}
G_{nm}^{-1}(\omega)-G_{nm}^{-1\ast}(\omega)=-2i\omega\Gamma(\delta_{n1}+\delta_{nN})\delta_{nm}. 
\label{a5}
\end{eqnarray}
Multiplying by $G(\omega)$ on the left and $G^\ast(\omega)$ on the right and using the symmetry of 
$G_{nm}(\omega)$ we obtain the Schwinger identity \cite{Wang45}
\begin{eqnarray}
G_{nm}(\omega)-G_{nm}^\ast(\omega)=2i\omega\Gamma
(G_{n1}(\omega)G_{1m}^\ast(\omega)+G_{nN}(\omega)G_{Nm}^\ast(\omega));
\label{a6}
\end{eqnarray}
the Schwinger identity in (\ref{a6}) is used later in Appendices \ref{current} and \ref{cgf}  in deriving the 
heat current and the cumulant generating function.
%%%%%%%%%%%%%%%%%%%%%%%%%%%%%%%%%%%%%%%%%%%%%%%%%%%%%%%%%
%%%%%%%%%%%%%%%%%%%%%%%%%%%%%%%%%%%%%%%%%%%%%%%%%%%%%%%%%
%%%%%%%%%%%%%%%%%%%%%%%%%%%%%%%%%%%%%%%%%%%%%%%%%%%%%%%%%
\subsubsection{Ordered chain - equation of motion method}
%%%%%%%%%%%%%%%%%%%%%%%%%%%%%%%%%%%%%%%%%%%%%%%%%%%%%%%%%
%%%%%%%%%%%%%%%%%%%%%%%%%%%%%%%%%%%%%%%%%%%%%%%%%%%%%%%%%
%%%%%%%%%%%%%%%%%%%%%%%%%%%%%%%%%%%%%%%%%%%%%%%%%%%%%%%%%
In the ordered case for $m_n=m$ and $\kappa_n=\kappa$ the Green's functions
$G_{n1}(\omega)$ and $G_{nN}(\omega)$ are easily determined by an equation of motion method. Alternatively,
one can employ a determinantal scheme noting that the determinant of $G^{-1}(\omega)$ for
$\tilde\Omega_1=\tilde\Omega_N=\Omega$ is given by $\kappa^NU_N(\Omega/2\kappa)$,
where $U_N(x)$ is the Chebychev polynomial of the second kind; 
$U_N(\Omega/2\kappa)=\sin p(N+1)/\sin p, \Omega=2\kappa \cos p$ \cite{Lebedev72}.

Addressing the equations of motion, which are of the linear difference form,
\begin{eqnarray}
&&\Omega\tilde u_n(\omega)=\kappa(\tilde u_{n+1}(\omega)+\tilde u_{n-1}(\omega)),
\label{a7}
\\
&&\Omega\tilde u_1(\omega)=i\Gamma\omega\tilde u_1(\omega)+\kappa\tilde u_2(\omega)+\tilde\xi_1(\omega),
\label{a8}
\\
&&\Omega\tilde u_N(\omega)=i\Gamma\omega\tilde u_N(\omega)+\kappa\tilde u_{N-1}(\omega)+\tilde\xi_2(\omega),
\label{a9}
\end{eqnarray}
and using the plane wave ansatz  $\tilde u_n=A\exp(ipn)+B\exp(-ipn)$ equation (\ref{a7}) yields 
$\Omega=2\kappa\cos p=\kappa(\exp(ip)+\exp(-ip))$. Inserting $\Omega$ in (\ref{a8}) and
(\ref{a9}) we obtain for the determination of $A$ and $B$ the matrix equation
\begin{eqnarray}
\left(\begin{array}{cc}
\kappa-i\Gamma\omega e^{ip} & \kappa-i\Gamma\omega e^{-ip}\\(\kappa e^{ip}-i\Gamma\omega)e^{ipN}
&(\kappa e^{-ip}-i\Gamma\omega)e^{-ipN}\end{array}\right)\left(\begin{array}{c}A\\B\end{array}\right)=
\left(\begin{array}{c}\xi_1 \\\xi_2\end{array}\right),
\label{a10}
\end{eqnarray}
which readily yields $A$ and $B$ and thus $\tilde u_n(\omega)$ as a function of $\tilde\xi_1(\omega)$ 
and $\tilde\xi_2(\omega)$. From (\ref{sol}) in Sec. \ref{model} we obtain the Green's functions
\begin{eqnarray}
&&G_{n1}(\omega)=\frac{\kappa\sin p(N+1-n)-i\Gamma\omega\sin p(N-n)}
{\kappa^2\sin p(N+1)-2i\kappa\Gamma\omega\sin pN-(\Gamma\omega)^2\sin p(N-1)},\label{g1}\\
&&G_{nN}(\omega)=\frac{\kappa\sin pn-i\Gamma\omega\sin p(n-1)}{\kappa^2\sin p(N+1)-2i\kappa\Gamma\omega\sin pN
-(\Gamma\omega)^2\sin p(N-1)},
\label{a11}
\end{eqnarray}
and in particular the end-to-end Green's function
\begin{eqnarray}
G_{1N}(\omega)=\frac{\kappa\sin p}{\kappa^2\sin p(N+1)-2i\kappa\Gamma\omega\sin pN-
(\Gamma\omega)^2\sin p(N-1)},
\label{a12}
\end{eqnarray}
i.e., the expression (\ref{green1}).
%%%%%%%%%%%%%%%%%%%%%%%%%%%%%%%%%%%%%%%%%%%%%%%%%%%%%%%%%
%%%%%%%%%%%%%%%%%%%%%%%%%%%%%%%%%%%%%%%%%%%%%%%%%%%%%%%%%
%%%%%%%%%%%%%%%%%%%%%%%%%%%%%%%%%%%%%%%%%%%%%%%%%%%%%%%%%
\subsubsection{Disordered chain - in terms of the transfer matrix}
%%%%%%%%%%%%%%%%%%%%%%%%%%%%%%%%%%%%%%%%%%%%%%%%%%%%%%%%%
%%%%%%%%%%%%%%%%%%%%%%%%%%%%%%%%%%%%%%%%%%%%%%%%%%%%%%%%%
%%%%%%%%%%%%%%%%%%%%%%%%%%%%%%%%%%%%%%%%%%%%%%%%%%%%%%%%%
In the disordered case we consider the equations of motion
\begin{eqnarray}
&&\Omega_n\tilde u_n(\omega)=\kappa_n\tilde u_{n+1}(\omega)+\kappa_{n-1}\tilde u_{n-1}(\omega),
\label{a13}
\\
&&\Omega_1\tilde u_1(\omega)=i\Gamma\omega\tilde u_1(\omega)+\kappa_1\tilde u_2(\omega)+\tilde\xi_1(\omega),
\label{a14}
\\
&&\Omega_N\tilde u_N(\omega)=i\Gamma\omega\tilde u_N(\omega)+\kappa_{N-1}\tilde u_{N-1}(\omega)+\tilde\xi_2(\omega).
\label{a15}
\end{eqnarray}
The bulk equation of motion (\ref{a13}) can be expressed in terms of a transfer matrix $T_n$
according to 
\begin{eqnarray}
\left(\begin{array}{c}\tilde u_{n+1}\\\tilde u_n\end{array}\right)=
\left(\begin{array}{cc}\Omega_n/\kappa_n &-\kappa_{n-1}/\kappa_n\\1 & 0\end{array}\right)
\left(\begin{array}{c}\tilde u_n\\\tilde u_{n-1}\end{array}\right)=T_n
\left(\begin{array}{c}\tilde u_n\\\tilde u_{n-1}\end{array}\right),
\label{a16}
\end{eqnarray}
and we have by successive applications
\begin{eqnarray}
\left(\begin{array}{c}\tilde u_N\\\tilde u_{N-1}\end{array}\right)=T_{N-1}T_{N-2}\cdots T_2
\left(\begin{array}{c}\tilde u_2\\\tilde u_1\end{array}\right).
\label{a17}
\end{eqnarray}
Likewise, from (\ref{a14}) and (\ref{a15}) we obtain
\begin{eqnarray}
\left(\begin{array}{c}\tilde u_2 \\\tilde u_1\end{array}\right)&=&
\left(\begin{array}{cc}(\Omega_1-i\omega\Gamma)/\kappa_1 & -1/\kappa_1 \\1 & 0\end{array}\right)
\left(\begin{array}{c}\tilde u_1 \\\tilde \xi_1\end{array}\right),
\label{a18}
\\
\nonumber
\\
\left(\begin{array}{c}\tilde u_N \\\tilde u_{N-1}\end{array}\right)&=&
\left(\begin{array}{cc}1 & 0 \\
(\Omega_N-i\omega\Gamma)/\kappa_{N-1} & -1/\kappa_{N-1}\end{array}\right)
\left(\begin{array}{c}\tilde u_N \\\tilde \xi_2\end{array}\right).
\label{a19}
\end{eqnarray}
Inserting (\ref{a18}) and (\ref{a19}) in (\ref{a17}) and using $B$ in (\ref{bom}) we have
\begin{eqnarray}
\left(\begin{array}{c}\tilde u_N \\\tilde \xi_2\end{array}\right)=
\tilde T\left(\begin{array}{c}\tilde u_1 \\\tilde \xi_1\end{array}\right),
\label{a20}
\end{eqnarray}
where
\begin{eqnarray}
\tilde T=\frac{1}{\kappa_0}\left(\begin{array}{cc}0 &1 \\\kappa_N & -i\Gamma\omega\end{array}\right)
B\left(\begin{array}{cc}\kappa_0 & 0 \\i\Gamma\omega & 1\end{array}\right).
\label{a21}
\end{eqnarray}
Expanding (\ref{a21}), using $\det\tilde T=-1$,  and comparing with (\ref{sol}) for $n=1$ and $n=N$, i.e.,
\begin{eqnarray}
&&\tilde u_1=G_{11}\tilde \xi_1+G_{1N}\tilde \xi_2,\label{u1}\\&&\tilde u_N=G_{N1}\tilde\xi_1+G_{NN}\tilde\xi_2,
\label{a22}
\end{eqnarray}
we infer the end-to-end Green's function
\begin{eqnarray}
&&G_{1N}=G_{N1}=1/\tilde T_{21}.
\label{a24}
\end{eqnarray}
From (\ref{a21}) we have
\begin{eqnarray}
\tilde T_{21}=\frac{1}{\kappa_0}(\kappa_0\kappa_NB_{11}
+i\Gamma\omega(\kappa_NB_{12}-\kappa_0B_{21})+(\Gamma\omega)^2B_{22}).
\label{a29}
\end{eqnarray}
Finally, from (\ref{a24}) and (\ref{a29}) we obtain (\ref{green2}), i.e.,
\begin{eqnarray}
G_{1N}(\omega)=\frac{\kappa_0}{\kappa_0\kappa_NB_{11}(\omega)+i\Gamma\omega
(\kappa_NB_{12}(\omega)-\kappa_0B_{21}(\omega))+(\Gamma\omega)^2B_{22}(\omega)}.
\label{a31}
\end{eqnarray}
%
%%%%%%%%%%%%%%%%%%%%%%%%%%%%%%%%%%%%%%%%%%%%%%%%%%%%%%%%%
%%%%%%%%%%%%%%%%%%%%%%%%%%%%%%%%%%%%%%%%%%%%%%%%%%%%%%%%%
%%%%%%%%%%%%%%%%%%%%%%%%%%%%%%%%%%%%%%%%%%%%%%%%%%%%%%%%%
\subsubsection{Ordered chain - special case of disordered chain}
%%%%%%%%%%%%%%%%%%%%%%%%%%%%%%%%%%%%%%%%%%%%%%%%%%%%%%%%%
%%%%%%%%%%%%%%%%%%%%%%%%%%%%%%%%%%%%%%%%%%%%%%%%%%%%%%%%%
%%%%%%%%%%%%%%%%%%%%%%%%%%%%%%%%%%%%%%%%%%%%%%%%%%%%%%%%%
In the ordered case for $m_n=m$ and $\kappa_n=\kappa$ the transfer matrix $T_n=T$ is independent 
of the site index $n$. We have
\begin{eqnarray}
T=\left(\begin{array}{cc}\Omega/\kappa & -1 \\1 & 0\end{array}\right).
\label{a32}
\end{eqnarray}
Setting $\Omega=2\kappa\cos p$ the matrix $T$ has the eigenvalues $\exp(\pm ip)$ forming the diagonal
matrix $D$ with matrix elements $\exp(\pm ip)$. Denoting the similarity transformation by $S$ we have $T=SDS^{-1}$ and thus
$T^n=SD^nS^{-1}$, where $D^n$ has the diagonal elements $\exp(\pm ipn)$. Finally, the similarity transformation $S$
has to be determined. However, a more direct way is again to apply the plane wave ansazt 
$\tilde u_n=A\exp(ipn)+B\exp(-ipn)$ to $\tilde u_1$ and $\tilde u_2$
and subsequently determine $A$ and $B$. We obtain in matrix form
\begin{eqnarray}
\left(\begin{array}{c}\tilde u_{n+1} \\\tilde u_n\end{array}\right)=\frac{1}{\sin p}
\left(\begin{array}{cc}\sin pn & -\sin p(n-1) \\\sin p(n-1) & -\sin p(n-2)\end{array}\right)
\left(\begin{array}{c}\tilde u_2 \\\tilde u_1\end{array}\right),
\label{a34}
\end{eqnarray}
and we  infer from (\ref{eqmo3}) the matrix product
\begin{eqnarray}
T^q=\frac{1}{\sin p}\left(\begin{array}{cc}\sin p(q+1) & -\sin pq \\\sin pq & -\sin p(q-1)\end{array}\right).
\label{a35}
\end{eqnarray}
We note that $T$ is in accordance with (\ref{trans1}) and that we have the group property $T^nT^m=T^{n+m}$.
%%%%%%%%%%%%%%%%%%%%%%%%%%%%%%%%%%%%%%%%%%%%%%%%%%%%%%%%%
%%%%%%%%%%%%%%%%%%%%%%%%%%%%%%%%%%%%%%%%%%%%%%%%%%%%%%%%%
%%%%%%%%%%%%%%%%%%%%%%%%%%%%%%%%%%%%%%%%%%%%%%%%%%%%%%%%%
\subsection{\label{liapunovapp}Liapunov exponent}
%%%%%%%%%%%%%%%%%%%%%%%%%%%%%%%%%%%%%%%%%%%%%%%%%%%%%%%%%
%%%%%%%%%%%%%%%%%%%%%%%%%%%%%%%%%%%%%%%%%%%%%%%%%%%%%%%%%
%%%%%%%%%%%%%%%%%%%%%%%%%%%%%%%%%%%%%%%%%%%%%%%%%%%%%%%%%
The evaluation of the Liapunov exponent is given by
$\gamma=\langle\epsilon\rangle=\int d\epsilon \epsilon P_0(\epsilon)/\int d\epsilon P_0(\epsilon)$ where 
$P_0(\epsilon)$ is the solution of the Fokker-Planck equation (\ref{fp}) and (\ref{pc}), i.e.,
\begin{eqnarray}
\Delta P_0'+a'P_0=C.
\label{b1}
\end{eqnarray}
Here a prime denotes a derivative with respect to $\epsilon$ and we have introduced the notation
$\Delta=(m\omega^2/\langle\kappa\rangle^2)^2\langle\delta\tilde\kappa^2\rangle$ and 
$a'(\epsilon)=2(\epsilon^2+(m/\langle\kappa\rangle)\omega^2)$; $C$ is an integration constant.

\begin{figure}
\includegraphics[width=0.8\hsize]{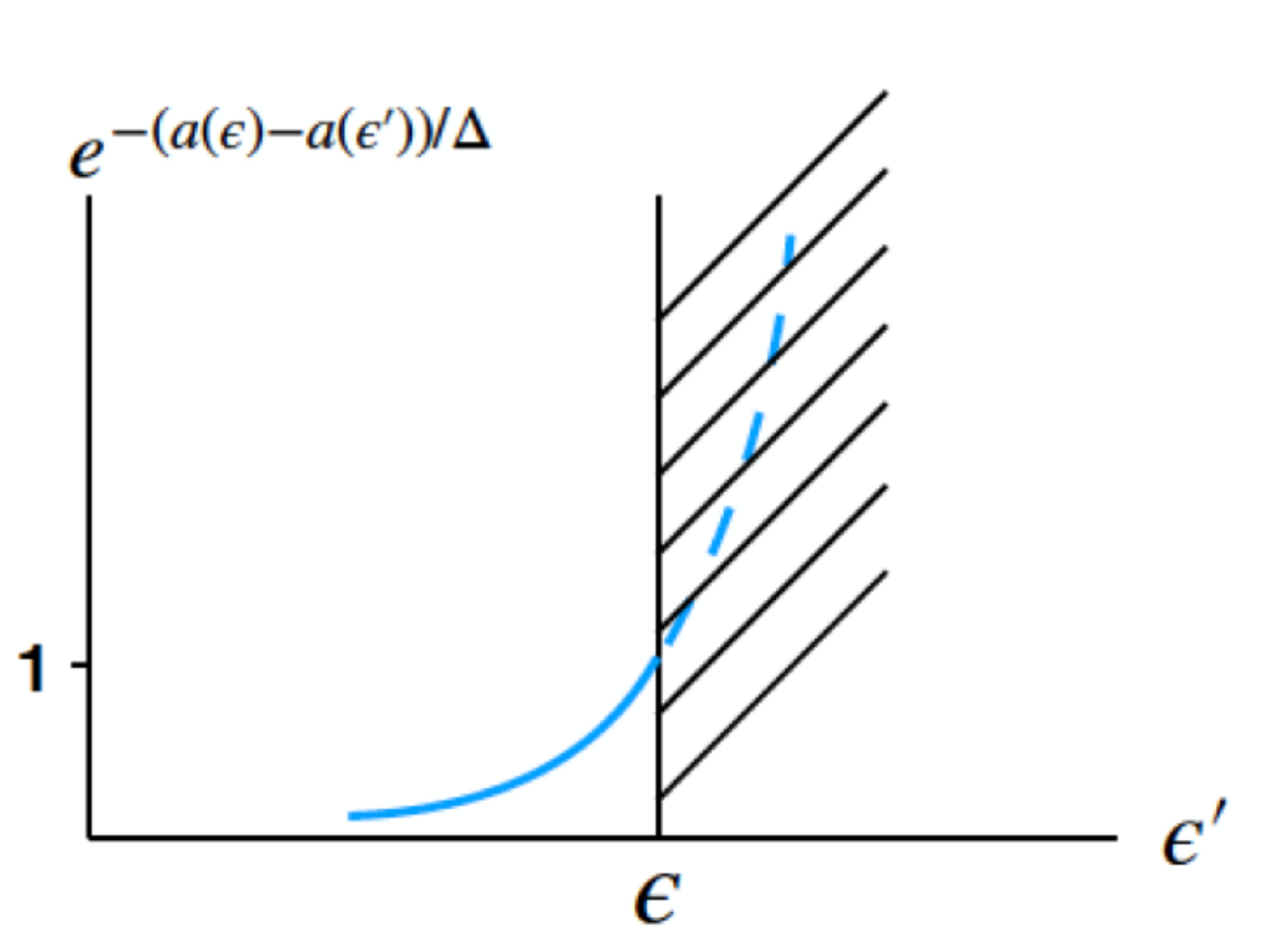}
\caption{This is a plot of the integrand $\exp(-(a(\epsilon)-a(\epsilon'))/\Delta)$ as a function of $\epsilon'$ 
in the expression (\ref{b3}) for the stationary distribution $P_0(\epsilon)$.}
\label{fig7}
\end{figure}

Assuming regularity in $\Delta$ and setting $P_0=P_0^{(1)}+\Delta P_0^{(2)}$ we obtain to leading order
\begin{eqnarray}
P_0=C\Bigg[\frac{1}{a'}+\Delta\frac{a''}{(a')^3}\Bigg].
\label{b2}
\end{eqnarray}
%t
This result can be justified by a steepest descent analysis. A particular solution of (\ref{b1}) has the form 
\begin{eqnarray}
P_0(\epsilon)=\frac{C}{\Delta}\int_0^\epsilon d\epsilon'\exp(-(a(\epsilon)-a(\epsilon'))/\Delta).
\label{b3}
\end{eqnarray}
In the exponent $a(\epsilon)=2(\epsilon^3/3+(m/\langle\kappa\rangle)\omega^2\epsilon)$ is a monotonically increasing function
passing through the origin. A plot of $\exp(-(a(\epsilon)-a(\epsilon'))/\Delta)$ is schematically
depicted in Fig.~\ref{fig7}. For small $\Delta$ the exponential function rises steeply and the main
contribution to $P(\epsilon)$ arises from the region $\epsilon'\le\epsilon$. To leading order a steepest descent
argument yields $P_0^{(1)}=C/a'$. The next term in the asymptotic expansion is obtained by expanding
$a(\epsilon')$, i.e., $a(\epsilon')=a(\epsilon)+a'(\epsilon)(\epsilon'-\epsilon)+(1/2)a''(\epsilon)(\epsilon'-\epsilon)^2$.
A straightforward calculation then yields $P_0^{(2)}=Ca''/a'^3$. In conclusion, the  expansion
in (\ref{stat}) is  an asymptotic expansion in $\Delta$, i.e., the leading correction to the steepest descent term.
%%%%%%%%%%%%%%%%%%%%%%%%%%%%%%%%%%%%%%%%%%%%%%%%%%%%%%%%%
%%%%%%%%%%%%%%%%%%%%%%%%%%%%%%%%%%%%%%%%%%%%%%%%%%%%%%%%%
%%%%%%%%%%%%%%%%%%%%%%%%%%%%%%%%%%%%%%%%%%%%%%%%%%%%%%%%%
\subsection{\label{current}Heat current}
%%%%%%%%%%%%%%%%%%%%%%%%%%%%%%%%%%%%%%%%%%%%%%%%%%%%%%%%%
%%%%%%%%%%%%%%%%%%%%%%%%%%%%%%%%%%%%%%%%%%%%%%%%%%%%%%%%%
%%%%%%%%%%%%%%%%%%%%%%%%%%%%%%%%%%%%%%%%%%%%%%%%%%%%%%%%%
Focussing on the heat reservoir at temperature $T_1$ at the site $n=1$, the integrated heat flux is obtained from
(\ref{qr1}), i.e.,
\begin{eqnarray}
Q(t)=\int^t dt'F_1(t')\dot u_1(t').
\label{c1}
\end{eqnarray}
Inserting $F_1=-\Gamma\dot u_1+\xi_1$  and $u_1$ from (\ref{sol}) we obtain in Fourier space
\begin{eqnarray}
Q(t)=\int\frac{d\omega}{2\pi}\frac{d\omega'}{2\pi}F(\omega-\omega',t)
\left(\begin{array}{cc}\tilde\xi_1(\omega) & \tilde\xi_2(\omega)\end{array}\right)M(\omega,\omega')
\left(\begin{array}{c}\tilde\xi_1(\omega') \\\tilde\xi_2(\omega')\end{array}\right),
\label{c2}
\end{eqnarray}
where the matrix elements of $M$ are given by
\begin{eqnarray}
&&M_{11}(\omega,\omega')=-\omega\omega'\Gamma G_{11}(\omega)G_{11}(-\omega')
+\frac{1}{2}\bigg(-i\omega G_{11}(\omega)+i\omega' G_{11}(-\omega')\bigg),
\label{c3}
\\
&&M_{22}(\omega,\omega')=-\omega\omega'\Gamma G_{1N}(\omega)G_{1N}(-\omega'),
\label{c4}
\\
&&M_{12}(\omega,\omega')=-\omega\omega'\Gamma G_{11}(\omega)G_{1N}(-\omega')
+\frac{1}{2}i\omega' G_{1N}(-\omega'),
\label{c5}
\\
&&M_{21}(\omega,\omega')=-\omega\omega'\Gamma
G_{1N}(\omega)G_{11}(-\omega')-\frac{1}{2}i\omega G_{1N}(\omega).
\label{c6}
\end{eqnarray}
The function $F(\omega,t)=2\sin(\omega t/2)\exp(-i\omega t/2)/\omega$. Moreover, 
$F(0,t)=t$ and $|F(\omega,t|^2=2\pi t\delta(\omega)$ for large $t$.
Finally, using the noise correlations (\ref{noi1}) and (\ref{noi2}) and the Schwinger identity (\ref{a6}), 
we obtain the mean heat flux $J=\langle Q(t)\rangle/t$ in (\ref{heat1}), i.e.,
\begin{eqnarray}
J=2(T_1-T_2)\int\frac{d\omega}{2\pi}(\omega\Gamma)^2|G_{1N}(\omega)|^2.
\label{c7}
\end{eqnarray}
%
%%%%%%%%%%%%%%%%%%%%%%%%%%%%%%%%%%%%%%%%%%%%%%%%%%%%%%%%%
%%%%%%%%%%%%%%%%%%%%%%%%%%%%%%%%%%%%%%%%%%%%%%%%%%%%%%%%%
%%%%%%%%%%%%%%%%%%%%%%%%%%%%%%%%%%%%%%%%%%%%%%%%%%%%%%%%%
\subsection{\label{cgf}Large deviation function}
%%%%%%%%%%%%%%%%%%%%%%%%%%%%%%%%%%%%%%%%%%%%%%%%%%%%%%%%%
%%%%%%%%%%%%%%%%%%%%%%%%%%%%%%%%%%%%%%%%%%%%%%%%%%%%%%%%%
%%%%%%%%%%%%%%%%%%%%%%%%%%%%%%%%%%%%%%%%%%%%%%%%%%%%%%%%%
The large deviation function function is defined according to
\begin{eqnarray}
\mu(\lambda)=\lim_{t\rightarrow\infty}\frac{1}{t}\ln\langle\exp(\lambda Q(t))\rangle,
\label{d1}
\end{eqnarray}
where $Q(t)$ is given by (\ref{c2}). Inserting $Q(t)$, using the noise distribution
\begin{eqnarray}
P(\xi_1,\xi_2)\propto\exp\Bigg[-\frac{1}{2}\int\frac{d\omega}{2\pi}\frac{d\omega'}{2\pi}
\left(\begin{array}{cc}\xi_1(\omega)&\xi_2(\omega)\end{array}\right)\Delta^{-1}(\omega-\omega')
\left(\begin{array}{c}\xi_1(\omega')\\ \xi_2(\omega')\end{array}\right)\Bigg], 
\label{d2}
\end{eqnarray}
where the inverse noise matrix is
\begin{eqnarray}
\Delta^{-1}(\omega-\omega')=
\left(\begin{array}{cc}\Delta^{-1}_1 & 0 \\0 & \Delta^{-1}_2\end{array}
\right)\delta(\omega-\omega'), 
\label{d3}
\end{eqnarray}
with $\Delta_1=2\Gamma T_1$ and $\Delta_2=2\Gamma T_2$,
and using the matrix identity \cite{Zinn-Justin89}
\begin{eqnarray}
\langle\exp(-(1/2\tilde\xi B\xi)\rangle=\exp(\rm{Tr}\ln(I+\Delta B)), 
\label{d4}
\end{eqnarray}
we obtain formally
\begin{eqnarray}
\mu(\lambda)=-\frac{1}{2t}\rm{Tr}\ln(I+2\lambda F\Delta M);
\label{d5}
\end{eqnarray}
we note that formula (\ref{d4}) follows from  (1.5) in Ref. \cite{Zinn-Justin89}
setting $b_i=0$ and using $\text{det} A=\exp(\text{Tr} \log A)$.

Using the properties of $F$, the limits for $\omega=\omega'$
\begin{eqnarray}
&&M_{11}(\omega,\omega)=-M_{22}(\omega,\omega)=\omega^2\Gamma |G_{1N}(\omega)|^2,
\label{d6}
\\
&&M_{12}(\omega,\omega)=M_{21}(\omega,\omega)^\ast=-\omega^2\Gamma G_{11}(\omega)
G_{1N}(\omega)^\ast+(1/2)i\omega G_{1N}(\omega)^\ast, 
\label{d7}
\end{eqnarray}
the Schwinger identity (\ref{a6}), and diagonalising $\Delta M$, we obtain the eigenvalue equation
for the eigenvalues  $\alpha_1(\omega)$ and $\alpha_2(\omega)$ 
\begin{eqnarray}
\alpha^2-2\Gamma\alpha(T_1M_{11}+T_2M_{22})+4\Gamma^2 T_1T_2(M_{11}M_{22}-M_{12}M_{21})=0.
\label{d8}
\end{eqnarray}
For $\mu$ we then obtain
\begin{eqnarray}
\mu(\lambda)=-\frac{1}{2}\int\frac{d\omega}{2\pi}\ln(1-2\lambda(\alpha_1+\alpha_2)+4\lambda^2\alpha_1\alpha_2),
\label{d9}
\end{eqnarray}
or reduced further the final result
\begin{eqnarray}
\mu(\lambda)=-\frac{1}{2}\int\frac{d\omega}{2\pi}\ln(1+4\omega^2\Gamma^2|G_{1N}(\omega)|^2f(\lambda)),
\label{d10}
\end{eqnarray}
where
\begin{eqnarray}
f(\lambda)=T_1T_2\lambda(1/T_1-1/T_2-\lambda).
\label{d11}
\end{eqnarray}
\newpage
%\bibliography{/Users/hansfogedby/Library/texmf/bibtex/bib/articles,books}

\newpage
\end{document}